\documentclass{aa}  
\usepackage[T1]{fontenc}
\usepackage{newtxtext,newtxmath} 
\usepackage{amsmath}
\usepackage{graphicx}
\usepackage{caption}
\usepackage{subcaption}
\usepackage{booktabs}
\usepackage{threeparttable}
\usepackage{tabularx}
\usepackage{longtable}
\usepackage{lscape} 
\usepackage{placeins}
\usepackage{makecell}
\usepackage[normalem]{ulem}

\usepackage{supertabular}
\usepackage{cuted}
\usepackage{xcolor}
\usepackage[table]{xcolor}
\usepackage{natbib}
\usepackage[colorlinks=true,
            urlcolor=blue,
            linkcolor=blue,
            citecolor=blue]{hyperref}       
\usepackage{xcolor}
\usepackage{ulem}
\usepackage{comment}
\usepackage[switch]{lineno}
\usepackage{lipsum}
\usepackage{xspace} 
\newcommand{\herx}{Her~X-1\xspace}
\newcommand{\suporb}{P_\textrm{sup}}
\newcommand{\dotsuporb}{\dot{P}_\textrm{sup}}
\newcommand{\oc}{$O-C$\xspace}
\newcommand{\batse}{\textit{CGRO}/\textrm{BATSE}\xspace}
\newcommand{\rxte}{\textit{RXTE}/\textrm{ASM}\xspace}
\newcommand{\swift}{\textit{Swift}/\textrm{BAT}\xspace}
\newcommand{\fermi}{\textit{Fermi}/\textrm{GBM}\xspace}
\newcommand{\maxi}{\textit{MAXI}/\textrm{GSC}\xspace}
\newcolumntype{Y}{>{\centering\arraybackslash}X}

\begin{document} 
\modulolinenumbers[1]

\title{Superorbital period evolution of Hercules X-1}
\author{Abhisek Tamang\inst{1,2}\thanks{E-mail: abhisek@rrimail.rri.res.in, abhisekt@iisc.ac.in}, Hemanth Manikantan\inst{3}, Kinjal Roy\inst{1}, Ajith Balu\inst{1} Biswajit Paul\inst{1}}
\institute{Raman Research Institute, C. V. Raman Avenue, Sadashivanagar, Bangalore, Karnataka - 560080, India\and
Joint Astronomy Programme, Indian Institute of Science, C. V. Raman Avenue, Bangalore, Karnataka - 560012, India\and
INAF - Istituto di Astrofisica e Planetologia Spaziali, Via del Fosso del Cavaliere, 100, 00133 Roma, Italy}

\date{Received MM DD, YYYY; accepted MM DD, YYYY}

\abstract
{Hercules X-1 (\herx) is one of the most well-studied X-ray binaries that exhibits superorbital modulation of its X-ray flux. The origin of the superorbital periodicity in \herx is still under debate between counter-precessing twisted-tilted accretion disc and free precession of the neutron star, with no clear consensus. The fluctuations in the superorbital periodicity in \herx with data obtained during 1970-1990 have been reported to be stochastic in nature with no long-term trend.}
{We investigated the long-term trend of the superorbital modulations in \herx with more recent and uninterrupted data during 1995-2025 to examine the presence of features that may favour or disfavour the proposed mechanisms. The main aim of this study is to test the stability of the superorbital period using data available from sensitive all-sky instruments.}
{We probed the variation in the superorbital period in \herx using long-term light curves from five X-ray monitors, namely \batse, \rxte, \swift, \fermi, and \maxi, along with data available in the literature, together spanning $\sim$50 yr. We employed multiple timing analysis tools to trace the long-term evolution of the superorbital period.}
{In the uninterrupted data of superorbital modulation for the last 30 years, the superorbital period in \herx is decreasing at the rate of $\sim -5 \times 10^{-5}$ s s$^{-1}$. The analysis of the \oc diagram showed that the superorbital period in \herx also has variabilities at shorter time scales on top of an overall decay. Hence, a simple quadratic function can not explain the variabilities observed in \oc curve. A positive correlation was found between the mean count rate and superorbital period obtained at different linear segments of the \oc curve.}
{The modulation of the superorbital period in \herx from a few to a few tens of superorbital cycles suggests that there are restoring mechanism in the system. We discuss the possible mechanisms responsible for the origin of the observed long-term decay and short-term variabilities.} 

\authorrunning{A. Tamang et. al}
\maketitle

\nolinenumbers

\section{Introduction}\label{sec:introduction}

Hercules X-1 (\herx) is a Low Mass X-ray Binary (LMXB) discovered in 1972 with the Uhuru satellite \citep{herx1_discovery}. Different distance values ranging from 5-7 kpc have been quoted for \herx, most recent being 7 kpc from Gaia data \citep{avakyan_xrbcats_2023}. It comprises a neutron star of mass $\sim 1.5\  M_\odot$ \citep{reynolds_new_1997} and a companion star, HZ Her, of spectral type A/F \citep{crampton_spectroscopic_1974} with a mass $\sim 2.2\ M_\odot$ \citep{deeter_pulse-timing_1981}. \herx is a pulsar with a 1.24 s spin period \citep{herx1_discovery}. The neutron star and the companion orbit each other about the common centre of mass with an orbital period of $1.7$ d \citep{middleditch_studies_1976} in an almost circular orbit with an eccentricity $\sim 0.0004$ \citep{staubert_updating_2009}. The orbital period is decaying at a modest rate of $\sim -4.85 \times 10^{-11}$ s s$^{-1}$ \citep{staubert_updating_2009}. The binary plane is almost edge-on with an inclination angle of $i=85.5^\circ \pm 0.3^\circ$ \citep{leahy_geometry_2025}, which leads to obscuration of the neutron star for $\sim 5.5$ h during eclipse \citep{herx1_discovery}. Along with the spin period and orbital period, the X-ray flux modulation from \herx exhibits an additional long-term periodicity of $\sim 35$ d, known as the superorbital period ($\suporb$) \citep{herx1_discovery, katz_thirty-five-day_1973, herx1_suporb_discovery}. \herx is one among the few X-ray binaries that exhibit superorbital periods, alongside other well-known binaries like LMC X-4 (30.5 d) \citep{lang_read_1981}, SMC X-1 (varies between 40-65 d) \citep{wojdowski_quasiperiodic_1998, trowbridge2007tracking}, and SS433 (162 d) \citep{hjellming_analysis_1981, margon_observations_1984}.

Each 35 d cycle in \herx comprises a high-intensity main-on state and a relatively lower-intensity short-on state (Fig.~\ref{fig:fig_sample_joint_gaussian_fit_cycle158}) covering the superorbital phases approximately 0-0.31 ($\sim 11$d) and 0.56-0.76 ($\sim 7$d), respectively \citep{scott_rxte_asm_35d, leahy_ginga_2000}, assuming the turn-on phase as the superorbital phase zero. \herx occasionally undergoes a complete off-state known as the anomalous low (AL). From its discovery to today, there have been five ALs (AL1-AL5) \citep{staubert_inversion_2017}. \herx has entered a new AL state from early 2026\footnote{\url{https://maxi.riken.jp/star_data/J1657+353/J1657+353.html}}.

The origin of the superorbital period in sources like LMC X-4, SMC X-1, and SS 433 has usually been attributed to the precession of the tilted accretion disc; however, its origin in \herx is still under debate. There are two mechanisms that have been put forward to explain the superorbital modulation in \herx. The first one suggests the presence of a counter-precessing, twisted-tilted accretion disc around the neutron star, which systematically obscures the X-ray emitting region, resulting in the observed 35 d modulation in the light-curve \citep{gerend_optical_1976, deeter_pulse-timing_1981, scott_35_2000, leahy_modelling_2002, leahy_35-day_2021, leahy_geometry_2025}. Another explanation is a free precession of the neutron star \citep{brecher_precession_1975, herx1_ns_free_precession_Truemper1986, klochkov_integral_2008, herx1_ns_free_precession_Postnov2013, herx1_ns_free_precession_Heyl2024}. So far, most of the work related to the superorbital period done has been devoted to understand the nature of the intensity modulation of the 35 d period and scatter in the duration of the individual superorbital cycles.

The 35 d period in \herx has typically been regarded as stable, with variations viewed as statistical fluctuations around a mean value. Early studies showed that the superorbital period of \herx is strictly synchronised to its orbital motion, where the superorbital period randomly picks up a value from 20, 20.5, and 21 times orbital period with equal probability \citep{STA83}. \cite{baykal_1993} (hereafter B93) used the turn-on times of the main-on of about 20 yr between 1970 and 1990 as a tracer to estimate the evolution of the superorbital period. They found that the fluctuations in turn-on times are consistent with a white noise random process. \cite{herx1_ns_free_precession_Postnov2013} later observed that the changes in pulse profile can be explained by free precession of the neutron star, and the random variation seen in the superorbital period can be explained by variability in the free precession of the neutron star by $\sim 2.5\%$ every few years. Recent work by \cite{leahy_swiftbat_2020} employed a cross-correlation method to estimate the main-on peak times from \rxte and \swift, and found that the distribution of the superorbital cycle lengths is consistent with a gaussian, also indicating no clear long-term trend in the superorbital period. 

In this work, we study the evolution of the superorbital period of \herx using main-on times reported in the literature together with data from X-ray all-sky monitors, spanning an almost 30-year continuous observational baseline.
 The paper is arranged in the following manner. In Sect.~\ref{sec:observation}, we describe instruments and data preparation. In Sect.~\ref{sec:analysis}, we investigate the evolution of the superorbital period using two independent methods: the epoch folding search ($\S$\ref{sec:epoch_folding_search}) and the deviation of the observed main-on peak times from the calculated values ($\S$\ref{sec:Pdot_suporb_from_O_C}). We discuss the results in Sect.~\ref{sec:discussion} and summarise them in Sect.~\ref{sec:summary}.

\section{Instruments and data preparation}\label{sec:observation}
The superorbital period of \herx is $\sim35$ d long. To investigate properties of the superorbital intensity variations, data from all-sky monitors are most suitable. We have utilised data from five such all-sky X-ray monitors, namely the \textit{Compton Gamma Ray Observatory (CGRO)}/Burst and Transient Source Experiment (BATSE), \textit{Rossi X-ray Timing Explorer (RXTE)}/All-sky monitor (ASM), \textit{Neil Gehrels Swift Observatory (Swift)}/Burst Alert Telescope (BAT), \textit{Fermi Gamma-ray Space Telescope (Fermi)}/Gamma Burst Monitor (GBM), and \textit{Monitor of All-sky X-ray Image (MAXI)}/Gas Slit Camera (GSC). In addition, we have also used older superorbital period data available in the literature to extend the temporal coverage (B93 and \citealt{STA06}; hereafter STA06).

\batse\ aboard the CGRO \citep{batse_meegan} was operational from 1991 to 2000. It comprised eight NaI-based scintillation detectors positioned to monitor hard photons from nearly the entire sky. The instrument was sensitive in the 20 keV--2 MeV energy range, enabling continuous observations of both transient and persistent high-energy sources.

\rxte aboard the RXTE \citep{asm_levine, rxte_jahoda} is an all-sky monitor that operated from 1996 to 2011. It had three position-sensitive Xenon proportional counters coupled to three coded-aperture masks, sensitive in the energy band 1.5--12 keV. With a total effective collecting area of about 90 cm$^2$, the instrument was capable of surveying approximately 80 \% of the sky during each $\sim$90 m orbital dwell.

\swift aboard the \textit{Swift} observatory \citep{Swift_Gehrels_2004, bat_2005} is an all-sky monitor that has been operational since 2004. It has Cadmium Zinc Telluride (CZT) detectors coupled to a 2D coded-aperture mask sensitive in the energy band 15--50 keV. It has a total detector area of 5200 cm$^2$ with an instantaneous field of view of 1.4 std. 

\fermi aboard the \textit{Fermi Gamma-ray Space Telescope} \citep{meegan2009fermi} is an all-sky monitor that has been operational since 2008. \fermi has an array of twelve sodium iodide scintillators and two bismuth germanate scintillators that together work in the energy band 8 keV--40 MeV. 

\maxi onboard \textit{MAXI} \citep{maxi, gsc_maxi}, which is mounted on the International Space Station (ISS), is an all-sky monitor that has been operational since 2009. It consists of 12 position-sensitive gas proportional counters working in the energy band 2--30 keV. With an instantaneous field of view of $160^\circ \times 1^\circ.5$ (FWHM), it scans the whole sky in one ISS orbital period.

\rxte, \swift, and \maxi measure the total intensity in the respective energy bands of the instruments, but \batse and \fermi measure the pulsed intensity. However, in the present work, the light curves are used solely to determine the timing of the superorbital cycle. Therefore, even though the absolute flux values may differ, the pulse intensity light curves from \batse and \fermi are still useful to study the timing properties of the superorbital period in \herx.

\begin{table}
\renewcommand{\arraystretch}{1.1}
\normalsize
\caption{Observation duration of the five instruments.}
\begin{threeparttable}
\begin{tabularx}{0.49\textwidth}{*{2}{>{\centering\arraybackslash}X}}
\hline
\hline
Instrument& Obs. duration (MJD)     \\
\hline
\batse          &   49899$-$51690   \\
\rxte           &   50087$-$55923   \\
\swift          &   53415$-$61069   \\
\fermi          &   54697$-$61147   \\
\maxi           &   55057$-$61155   \\
\hline                                                          
\end{tabularx}
\end{threeparttable}
\label{tab:observation_duration}
\end{table}

\rxte\footnote{\url{https://heasarc.gsfc.nasa.gov/cgi-bin/W3Browse/w3browse.pl}}, \swift\footnote{\url{https://swift.gsfc.nasa.gov/results/transients/}}, and \maxi\footnote{\url{http://maxi.riken.jp/mxondem/}} long-term light curves were downloaded with a bin size equal to the orbital period of the respective satellites. \batse and \fermi light curves were available with a bin size of 1 d and 2 d, respectively\footnote{\url{https://gammaray.nsstc.nasa.gov/gbm/science/pulsars.html}}. The observation duration of the light curve from each instrument is given in Table~\ref{tab:observation_duration}. 

We excluded the AL3 data from \rxte and removed the orbital eclipses from the \rxte, \swift, and \maxi light curves to reduce count-rate variations arising from variable sampling of eclipse durations. These light curves were then rebinned to 1 d resolution. \batse and \fermi light curves were retained at their original bin sizes.

\section{Analysis}\label{sec:analysis}

\subsection{Epoch folding search}\label{sec:epoch_folding_search}

First, we investigated the temporal evolution of the superorbital period using the epoch folding search method. We derived a single 30 yr long light curve of 2 d bin size by combining light curves from all five instruments. To account for differences in flux levels among instruments, we first arranged the light curves in chronological order based on their start times. We then calculated a scale factor for the first two light curves using the ratio of their weighted mean count rates in the overlapping region and used it to combine them. The resulting light curve was subsequently combined with the light curve from the next instrument, following the same procedure. This process was repeated sequentially until all five light curves were merged into a single light curve spanning 30 yr.

\begin{figure}[ht]
  \centering
  \includegraphics[width=\columnwidth]{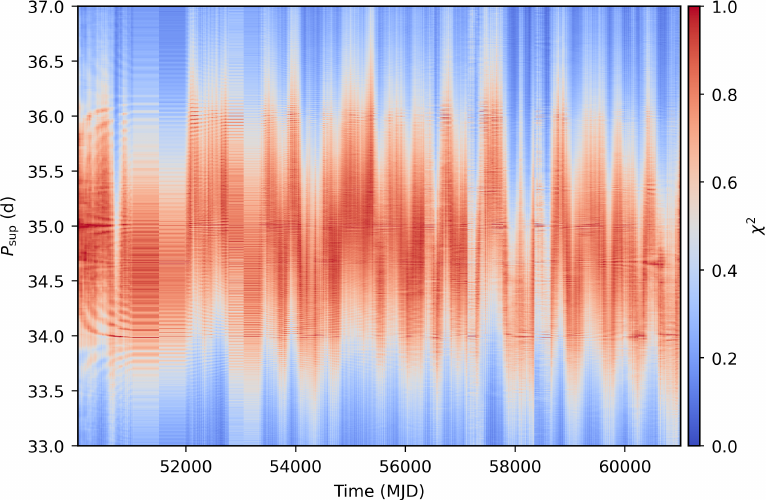}
  \caption{Periodogram of the superorbital period in \herx obtained using dynamic efsearch on the combined light curve. The Y-axis represents the superorbital period (d), the X-axis represents time (MJD), and the surface plot indicates the normalised $\chi^2$ values.} 
  \label{fig:const_Psup_evolution}
\end{figure}

We have used the epoch folding search tool, \texttt{efsearch}, available in the \texttt{XRONOS v6.0} timing analysis package to determine the superorbital period \citep{di_gesu_xronos_1992}. To probe the long-term evolution of the superorbital period, we searched for periods around 35 d inside sliding windows of length 280 d with a step size of 14 d applied to the combined light curve. The number of folding phase bins was set to 16, and the period search was performed with a resolution of 200 s over 10000 trial periods. Temporal evolution of the superorbital period obtained from the dynamic period search is shown in Fig.~\ref{fig:const_Psup_evolution}, where the colorbar (on the right) denotes the chi-squared ($\chi^2$) for each pair of superorbital period and time. The $\chi^2$ values in each sliding window were normalised to their respective maximum value. The dynamic period search shows numerous fluctuations throughout 30 yr of observation duration. Previous study reported that these fluctuations in the superorbital period in \herx are approximately Gaussian \citep{leahy_swiftbat_2020}. While in SMC X-1, the variation in the superorbital period is periodic in nature with a periodicity of $\sim 5$ yr \citep{clarkson_I, hu_superorbital_2013}.

\begin{figure}[ht]
  \centering
  \includegraphics[width=\columnwidth]{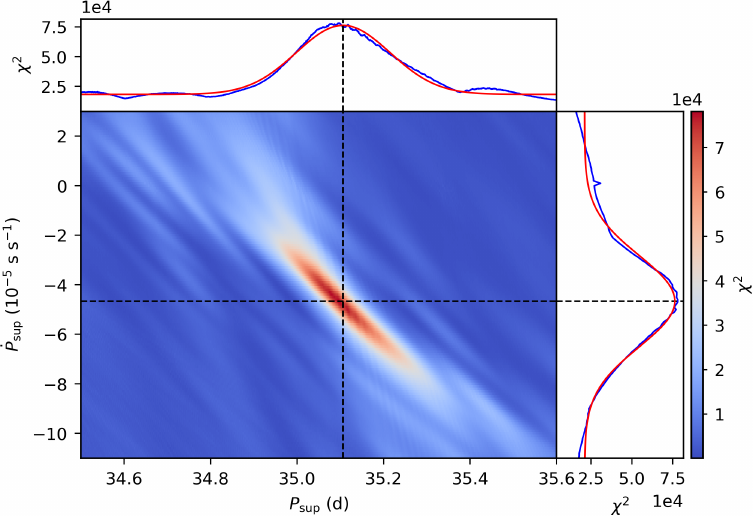}
  \caption{Surface plot of the $\chi^2-$ distribution as a function of the superorbital period and superorbital period derivative for \herx. Middle panel shows $\suporb$ on the X-axis for different values of $\dotsuporb$ on the Y-axis. The colour bar on the right represents the maximum $\chi^2$ corresponding to each ($\suporb, \dotsuporb$) pair. The top panel shows the variation of $\chi^2$ with $\suporb$ while the right panel shows the variation of $\chi^2$ with $\dotsuporb$. Two black lines are drawn at the best period and period derivative obtained from the gaussian fit (red) to the respective $\chi^2-$ distribution.}
  \label{fig:ppdot_chi_square_distribution}
\end{figure}

\begin{figure}[ht]
  \centering
  \includegraphics[width=0.9\columnwidth]{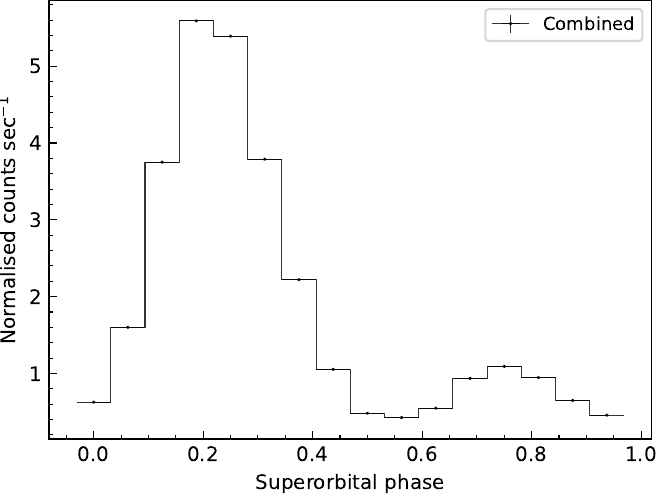}
  \caption{Superorbital profile of \herx created from the combined light curve using 16 bins. Superorbital period and superorbital period derivative used for folding are obtained from $\chi^2-$maximisation in $\suporb-\dotsuporb$ space created using epoch folding search (Fig.~\ref{fig:ppdot_chi_square_distribution}).}
  \label{fig:superorbital_profile}
\end{figure}

To further investigate the long-term evolution of the superorbital period, we ran \texttt{efsearch} on the combined light curve with trial periods around 35 d and period derivatives in the range $-3 \times 10^{-3}$ s s$^{-1}$ to $+3 \times 10^{-3}$ s s$^{-1}$ in steps of $10^{-6}$ s s$^{-1}$. The $\chi^2$ distribution in the period and period-derivative plane, as shown in Fig.~\ref{fig:ppdot_chi_square_distribution}, indicates the presence of a non-zero superorbital period derivative in \herx. The best superorbital period and its derivative obtained from $\chi^2$-maximisation are $\suporb=35.1(1)$ d and $\dotsuporb=-5(2) \times 10^{-5}$ s s$^{-1}$, respectively, referenced to the epoch 49900 MJD (Table~\ref{tab:efsearch_and_oc_analysis_table}). The quoted uncertainties represent the 1$\sigma$ confidence intervals, obtained by fitting gaussian functions to the corresponding $\chi^2$ distributions. The average superorbital profile obtained by folding the combined light curve using the obtained best-fit superorbital period and superorbital period derivative is shown in Fig.~\ref{fig:superorbital_profile}.

\begin{table*}[htp]
\centering
\renewcommand{\arraystretch}{1.12}
\normalsize
\caption{Superorbital periods and their derivatives obtained from epoch-folding and \oc analyses of the superorbital modulation data spanning the last 30 yr.}
\label{tab:combined_oc_table}
\begin{threeparttable}
\begin{tabularx}{0.92\textwidth}{>{\centering\arraybackslash}X
                                >{\centering\arraybackslash}X
                                >{\centering\arraybackslash}X
                                >{\centering\arraybackslash}X}
\toprule
Analysis  &
Observation duration &
$P_{\rm sup}$\tnote{a} &
$\dot{P}_{\rm sup}$\tnote{a} \\
& (MJD) & (d) &
($10^{-5}\,\mathrm{s\,s^{-1}}$) \\
\midrule
Efsearch & 49899--61155 & 35.1(1)   & -5(2) \\
\oc     & -do-         & 35.1(1) & -4.7(1) \\
\bottomrule
\end{tabularx}

\begin{tablenotes}
\item[a] Uncertainties are quoted at the 1$\sigma$ confidence level. Estimated superorbital period is referenced to the epoch 49900 MJD.
\end{tablenotes}
\end{threeparttable}
\label{tab:efsearch_and_oc_analysis_table}
\end{table*}

\subsection{\oc diagram from main-on peak timings}\label{sec:Pdot_suporb_from_O_C}

To examine the long-term evolution and observed superorbital period derivative in greater detail, we carried out an \oc analysis. The observed event arrival times are examined for deviations from an expected constant periodicity, with the resulting timing residuals revealing the long-term evolution of the period.

Superorbital profile of \herx is characterised by two stable features, a high-intensity main-on followed by a relatively fainter short-on with an off-state in between (Fig.~\ref{fig:superorbital_profile}). Although the length of the superorbital cycle varies about the mean value (B93, \citealt{leahy_swiftbat_2020}), no significant changes in the overall profile shape have been reported, except during complete off states, such as ALs. We therefore used the peak of the main-on as a tracer to estimate the evolution of the superorbital period, as the main-on state is prominent even in \batse and \fermi light curves. For a constant superorbital period, the expected time of arrival of the $n^\textrm{th}$ cycle main-on peak is given by:

\begin{equation}
    C_n = C_0 + n\suporb,
    \label{eq:linear_equation}
\end{equation}

where $\suporb$ is the superorbital period, $C_0$ and $C_n$ are expected times of arrival of the zeroth and $n^\textrm{th}$ cycle main-on peaks respectively. Deviations of the observed peak times from this relation indicate changes in the superorbital period. This technique was used to estimate the evolution of the orbital or superorbital period in various X-ray binaries, using characteristic events as tracers $-$ arrival times of orbital intensity minima for Cyg X-3 \citep{singh_cygx1_2002}, turn-on times of superorbital cycle \citep{paul_superorbital_2002} or mid-eclipse times in LMC X-4 \citep{lmcx4_orbPdot_Naik_2004}, mid-eclipse times in Cen X-3 and SMC X-1 \citep{cenx3_smcx1_orbPdot_raichur_2010} and peak flares times for GX 301-2 \citep{manikantan_gx301m2}.

\begin{figure}[ht]
  \centering
  \includegraphics[width=\columnwidth]{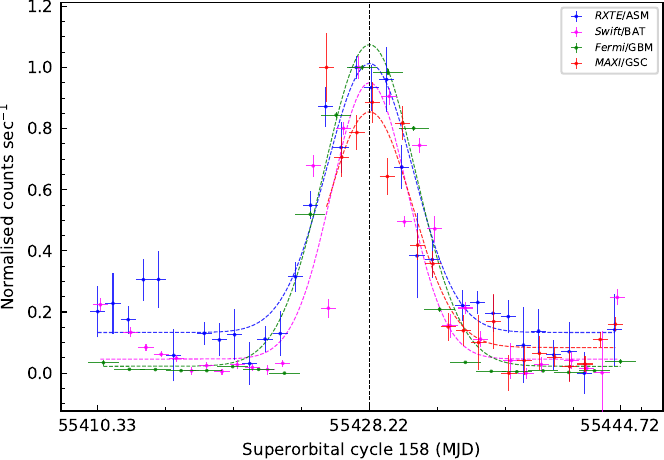}
  \caption{Sample joint gaussian-fit of the main-on profile for superorbital cycle 158, derived from the \rxte (blue), \swift (magenta), \fermi (green), and \maxi (red) light curves, respectively. Vertical line (black) is drawn at the main-on peak time obtained from joint fitting.}
  \label{fig:fig_sample_joint_gaussian_fit_cycle158}
\end{figure}

With the presence of a period derivative, the arrival time of the observed main-on peak time for the $n^\textrm{th}$ superorbital cycle can be written as

\begin{equation}
    O_n = C_0 + n. \suporb + \frac{n^2}{2!}.\suporb. \dotsuporb,
    \label{eq:quadratic_equation}
\end{equation}

where $\dotsuporb$ is the superorbital period derivative, and $O_n$ is the observed arrival time of the main-on peak of the $n^\textrm{th}$ superorbital cycle. 

In order to determine the observed main-on peak times from the long-term light curves, we modelled each main-on state with a gaussian function near the main-on peak of \herx. In cases where data from multiple missions overlapped, we jointly modelled the main-on state assuming a common main-on peak time with width and normalisation left free to vary. We also rejected all those cycles that had poor statistical quality or missing data points during main-on. A sample joint gaussian-fit of the main-on for the superorbital cycle 158 obtained from the four instruments (\rxte, blue; \swift, magenta; \fermi, green; and \maxi, red) is shown in Fig.~\ref{fig:fig_sample_joint_gaussian_fit_cycle158}. The time corresponding to the peak of the gaussian obtained from $n^{\mathrm{th}}$ superorbital cycle is the observed main-on peak time. 

\begin{figure}[htp]
  \centering
  \includegraphics[width=\columnwidth]{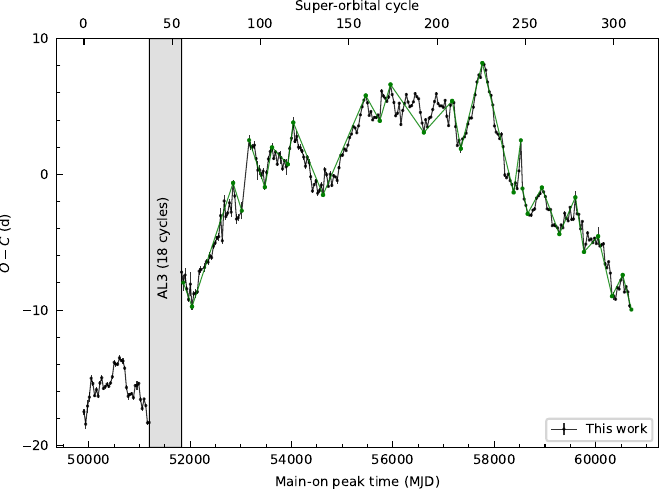}
  \caption{\oc\ diagram showing the deviation of the observed main-on peak times from the calculated main-on peak times obtained using the last 30 yr of superorbital modulation data of \herx. 24 Segments (green) represent the locally linear trends.}
  \label{fig:30yr_oc_variation}
\end{figure}

These main-on peak times were first fitted using Eq.~\eqref{eq:linear_equation} to estimate the superorbital period under the assumption of a constant period. The best-fit linear term was subtracted from the extracted main-on peak times to estimate \oc curve. The plot of combined \oc against the superorbital cycle numbers is shown in Fig.~\ref{fig:30yr_oc_variation}. An overall negative curvature can be seen in the \oc plot, which indicates a long-term decay of the superorbital period. The shaded region corresponds to the AL3 spanning approximately 18 superorbital cycles.

The observed main-on peak times were fitted with a quadratic function of the form (Eq.~\ref{eq:quadratic_equation}) to estimate the superorbital period and superorbital period derivative of 35.1(1) d and $-4.7(1) \times 10^{-5}\text{ s s}^{-1}$ with respect to the epoch 49900 MJD (Table~\ref{tab:efsearch_and_oc_analysis_table}). The errors are quoted at 1 $\sigma$ confidence level. The obtained superorbital period derivative is consistent with that derived from the $\chi^2-$ maximisation (Table~\ref{tab:efsearch_and_oc_analysis_table}) within the uncertainties. However, the formal 1$\sigma$ errors listed in Table~\ref{tab:efsearch_and_oc_analysis_table} should be treated with caution, as \oc curve shows clear deviations from a quadratic trend (Fig.~\ref{fig:30yr_oc_variation}), indicating that a simple quadratic model does not adequately describe the data.

\begin{figure}[htp]
  \centering
  \includegraphics[width=\columnwidth]{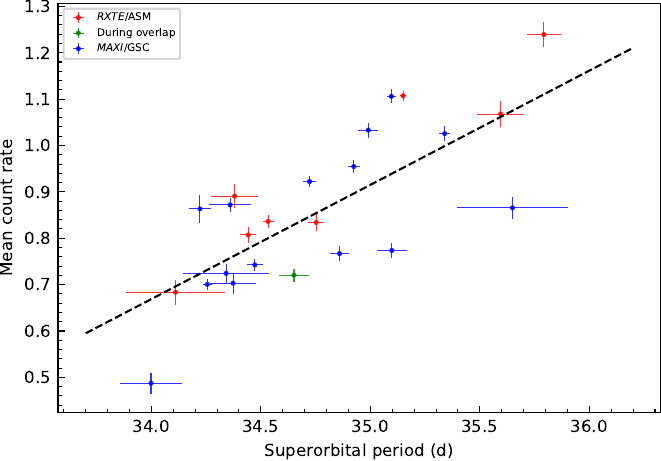}
  \caption{Correlation between the superorbital period at different epochs of the \oc and the corresponding mean count rates obtained from the \rxte light curve (red) and \maxi light curve (blue), where \maxi count rate is converted to equivalent \rxte count rate. Data in the overlapping region is marked in green.}
  \label{fig:Psup_cts_correlation}
\end{figure}

The \oc curve exhibits variability on multiple timescales. Additionally, excursions are present over the timescales of 10-38 superorbital cycles about the quadratic trend. Some of the prominent excursions are during MJDs $\sim$48550-49869, 49869-51191, 54136-55392, and 57500-58242. We estimated the superorbital periods in different smaller windows by selecting 24 locally linear trends from the \oc curve in the post-AL3, including the excursions (Fig.~\ref{fig:30yr_oc_variation}) using Ramer-Douglas-Peucker algorithm~\citep{ramer_iterative_1972, douglas_algorithms_1973}. A similar approach was previously employed to investigate the evolution of period in various sources such as XTE J1710-281 \citep{jain_eclipse_2022}, Vela X-1~\citep{liao_torque_2022}, and Cen X-3~\citep{liao_torque-dependent_2023}. During these excursions, the superorbital period changes by up to 5 \% between the maximum and the minimum values. Estimated superorbital periods at these 24 epochs and the corresponding mean count rates obtained from the \rxte and \maxi count rate converted to equivalent \rxte count rate, using the same procedure followed for combining the five light curves, showed a positive correlation (Fig~\ref{fig:Psup_cts_correlation}). The Pearson correlation coefficient (r) between the two quantities is 0.73, with a p-value $3.1 \times 10^{-5}$.

The five long-term light curves together cover approximately $30$ yr of superorbital cycles of \herx from $\sim1995 - 2025$. To investigate the period evolution over a longer baseline, we made use of the list of turn-on times from 1970--1990 (B93) and early \batse\ data, which are not publicly available (STA06). Since B93 and STA06 reported the superorbital cycle turn-on times, we calculated the corresponding main-on peak times to ensure consistency with the timing definition adopted in this work. The superorbital profile obtained from the long-term \herx light curves shows that the main-on peak occurs, on average, $\sim7$ d after the turn-on. We therefore added 7 d to the reported turn-on times to estimate the corresponding main-on peak times, providing nearly continuous coverage of superorbital modulation over $\sim$50 yr from 1970 to 2025. Estimated main-on peak times covering 50 yr of superorbital modulation of \herx are listed in Table~\ref{tab:mainon_peak_times}.

\begin{figure}[ht]
  \centering
  \includegraphics[width=\columnwidth]{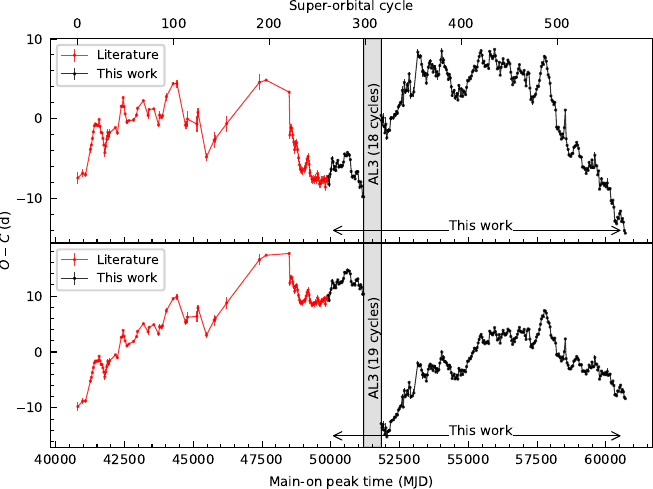}
  \caption{\oc diagram showing the deviation of the observed main-on peak times from the calculated values, constructed from 50 yr of superorbital modulation data, including both the present work (black) and literature data (red). The top (bottom) panels show diagrams obtained by assuming 18 (19) superorbital cycles elapsed during AL3. Blue arrows indicate the significant excursions.}
  \label{fig:50yr_oc_variation}
\end{figure}

The last observed turn-on times before the onset of the five ALs occurred at MJD 45469 (AL1; \citealt{delgado_al1_hz_1983, parmar_al1_an_1985}), MJD 49205 (AL2; \citealt{vrtilek_al2_multiwavelength_1994}), MJD 51224 (AL3; \citealt{parmar_al3_x-ray_1999, coburn_al3_the_2000, still_al3_rxte_2001}), MJD 52945 (AL4), and MJD 57302 (AL5; \citealt{staubert_inversion_2017}), with corresponding durations of 283, 103, 602, 212, and 70 d, respectively until the first turn-on after the end of ALs. During the AL3, which is by far the longest known low state of \herx, the exact number of superorbital cycles elapsed is not known. However, it has been reported that an extra cycle might have been completed \citep{staubert_two_2009}. Additionally, in our analysis, one more main-on peak time at the beginning of AL3 could not be constrained therefore, the total number of superorbital cycles elapsed during AL3 in two cases was 18 and 19. We constructed \oc\ diagrams for both cases (shown in Fig.~\ref{fig:50yr_oc_variation}), using superorbital periods of 34.862(1) d and 34.787(1) d, corresponding to 18 and 19 cycles during AL3, respectively.

Inspection of Fig.~\ref{fig:50yr_oc_variation} reveals that the long-term \oc curve is significantly different from a simple quadratic function (Eq.~\ref{eq:quadratic_equation}). Variability in the superorbital period is present throughout 50 yr of data.

\section{Discussion}\label{sec:discussion}

In this work, we investigated the long-term evolution of the superorbital period of the low-mass X-ray binary pulsar, \herx. For the first time, we report the presence of a non-zero period derivative in its superorbital period between the anomalous low states AL3 (1999) and AL6 (2026) using data from five X-ray monitor instruments. We confirmed the presence of this non-zero superorbital period derivative with two independent methods, yielding consistent results. Epoch folding search in superorbital period and superorbital period derivative space obtained $\dotsuporb = 5(2) \times 10^{-5}$ s s$^{-1}$ (Fig.~\ref{fig:ppdot_chi_square_distribution}), while \oc analysis gave $\dotsuporb = -4.7(1) \times 10^{-5}$ s s$^{-1}$ (Fig.~\ref{fig:30yr_oc_variation}).
 
In addition to the secular decay in $P_{\rm sup}$, the \oc curve reveals variability on multiple timescales, with excursions spanning 10-38 superorbital cycles and shorter-term fluctuations. Analysis of 24 locally linear trends indicates that these local variations in $\suporb$ can reach 5 \% (Fig.~\ref{fig:Psup_cts_correlation}). A positive correlation (r$=$0.73) between the superorbital period and X-ray count rates was also observed during these local fluctuations.

Using turn-on times from the literature, we also performed \oc analysis for a combined data set spanning 1970 to 2025 (Fig.~\ref{fig:50yr_oc_variation}). Though there is a clear superorbital period decay trend after AL3, described by a quadratic profile, it could not describe the overall 50 yr of \oc behaviour. Local fluctuations in superorbital period are ubiquitous throughout the 50 yr of observations, and these fluctuations are consistent with random-walk models in phase reported previously \citep{STA83, OGE85}. Additionally, distinct excursions spanning 10-38 superorbital cycles are present in 1995-2025 (Fig.~\ref{fig:30yr_oc_variation} and ~\ref{fig:50yr_oc_variation}), which have not been reported previously. The occurrence of these excursions (Figs.~\ref{fig:30yr_oc_variation} and Fig.~\ref{fig:50yr_oc_variation}) together with the positive correlation observed between the superorbital period and the average count rate (Fig.~\ref{fig:Psup_cts_correlation}) on these timescales suggest that radiation-driven restoring mechanisms such as X-ray irradiation of the accretion disc or the companion star may play some role in the variations of the superorbital period \citep{pringle_self-induced_1996, ogilvie_precessing_2001}.

Early studies of the superorbital modulation in \herx suggested a possible decay of the superorbital period, with  $\suporb/\dotsuporb\sim1625$ yr \citep{OGE85}. However, B93, with a longer observation duration, found no evidence for either a decreasing or an increasing trend in the superorbital period of \herx, with the observed fluctuations attributed to white noise. Similar results were obtained by \citet{leahy_swiftbat_2020}. After AL3, a characteristic time scale of $\suporb/\dotsuporb\sim1918$ yr (Table~\ref{tab:efsearch_and_oc_analysis_table}) has been obtained in our analysis.

In the following two sections, we discuss the possible origin of both small and large excursions, origin of superorbital period, and the observed secular decay post-AL3 in the light of disc precession and neutron star free precession models.

\subsection{Disc precession}\label{sec:disc_precession}

One of the widely accepted mechanisms for the origin of the 35 d periodicity in \herx is the precession of the accretion disc. The accretion disc precesses around the central neutron star while systematically obscuring the X-ray emission that gives rise to 35 d periodicity. Disc precession resulting from tidal interaction with the donor occurs if the mass ratio $<0.25 -0.33$ \citep{whitehurst_superhumps_1991} which is not the case for \herx. The most possible explanation for the precession is due to the non-axisymmetric radiation pressure from the central neutron star. When the central object illuminates the surface of an optically thick accretion disc non-uniformly, the backreaction of the emitted radiation causes tilting and warping of the disc, which eventually leads to precession \citep{pringle_self-induced_1996, maloney_radiation-driven_1996}. 

Radiation-driven warping of the accretion disc in Her~X-1 was first proposed by \citet{petterson_35_1977} and later developed theoretically by \citet{ogilvie_precessing_2001}. They showed that accretion discs can develop radiation-induced warps, with properties that depend on the binary separation and mass ratio. In their framework, the stability of these warps is determined by the range of the mass input radius, bounded by the circularisation and outer disc radii. However, their model does not have tidal torque and assumed a constant viscosity and accretion efficiency parameters. \citet{clarkson_I} using this model employed the dynamic power spectrum method on \rxte and \batse data and concluded that the periodic variation in the superorbital period of SMC X-1 arises from the interaction of warping modes. Building upon this work, \cite{clarkson_II} applied the same technique to study the X-ray binaries: LMC X-4, \herx, and Cyg X-2 using \rxte data. They concluded that among these three sources, LMC X-4 and \herx exhibit stable warp modes, while Cyg X-2 shows interactions of multiple warp modes. Their analysis further revealed that the AL3 in \herx was probably caused by changes in the inclination of the warped accretion disc, which leads to the complete obscuration of the central X-ray emitting region. In contrast to SMC X-1, the interaction of warp modes which can cause periodic variation in the superorbital period of \herx varies from $\sim 1-4$ yr time scales ($\sim 10-38$ superorbital cycles; Fig.~\ref{fig:50yr_oc_variation}), compared to $\sim5$ yr in SMC X-1 \citep{clarkson_I}. This demands extended continuous observations that were not available during these earlier studies.

\citet{ogilvie_precessing_2001} further showed that systems exhibiting perfectly steady superorbital periods, produced by stable, steadily precessing warped accretion discs, are unlikely. This implies that finite tidal torques from the donor and temporal variations in the mass accretion rate are generally unavoidable. In addition, interactions between higher-order warp modes can lead to quasi-periodic or aperiodic variations in the superorbital modulation, potentially contributing to long-term changes in superorbital period. \citet{kotze_characterizing_2012} further supported this interpretation by applying the dynamic power spectrum technique to a sample of 25 X-ray binaries, including \herx. Their analysis demonstrated that the superorbital variability in these systems is better described by time-dependent, interacting disc-warp modes rather than by strictly steady precession. Such evolving warp behaviour naturally accounts for the excursions observed in Fig.~\ref{fig:50yr_oc_variation}. 

\subsection{Neutron Star Precession}\label{sec:ns_free_precession}

{An alternative explanation for the 35 d cycle in \herx is a free precession of the neutron star. In this framework, the neutron star provides a stable 35 d clock, while the accretion disc precession becomes synchronised to this clock through strong coupling mechanisms \citep[and references therein]{shakura2021nature}. A slightly aspherical spinning neutron star can precess freely when its spin vector ($\vec{\Omega}$) is misaligned with a principal inertia axis ($I_n$), even in the absence of an external torque.

Observations of \herx with \textit{RXTE}/PCA between 1996$-$2005 revealed systematic changes in the 3$-$20 keV pulse profiles with superorbital phase, while retaining the shapes across different superorbital cycles \citep[and references therein]{staubert_variable_2013}. This behaviour was suggested to be a geometric effect of viewing a freely precessing neutron star \citep{herx1_ns_free_precession_Postnov2013}. \citet{shakura2021nature} showed that a freely precessing biaxial prolate neutron star ($I_3<I_2=I_1$) with a small ellipticity $\epsilon\sim10^{-6}$ and a spin-misalignment angle $\gamma\sim50^\circ$ (angle between $I_3$ and $\Omega$) can reproduce such superorbital phase-dependent pulse profile variations. By introducing a slight triaxial ($I_1< I_2 <I_3$) deformity to this model and assuming long intervals of almost constant precession period, \cite{kolesnikov2022evidence} extended this model to explain the \fermi measured spin frequency modulation with superorbital phase.

Recently, rotating vector model \citep{radhakrishnan1969magnetic} fits performed on X-ray polarisation measurements by \textit{IXPE} have constrained the neutron star crust spin ($\vec{\Omega}$) and magnetic dipole orientations (magnetic obliquity $\theta$ and $\phi_0$) of \herx \citep{herx1_ns_free_precession_Heyl2024}. The authors argued that the observed variation in $\theta$ between progressive superorbital phases, without an associated change in orientation of $\vec{\Omega}$, along with the fact that the external torque required to produce such a change exceeds that supplied by the disk, is indicative of freely precessing asymmetric neutron star crust ($\epsilon\sim10^{-7}$). However, they also reported a long-term drift in $\vec{\Omega}$ orientation at the same superorbital phase over longer time scales, which was attributed to residual accretion torque. Several different mechanisms that supply external torques on the neutron star crust have been discussed in \cite{staubert_variable_2013}.

Throughout the monitoring period, we observed several different types of variations in the superorbital period, superimposed on the secular trend. These variations in precession period over a few superorbital cycles are obvious from the Fig.~\ref{fig:30yr_oc_variation}. Occasional excursions over the timescales of 10-38 superorbital cycles seem to develop and subside gradually, hinting at the presence of some restoring mechanism (Fig.~\ref{fig:30yr_oc_variation} and~\ref{fig:50yr_oc_variation}). However, it is unlikely to be related to the internal damping due to coupling between neutron star layers, as the damping time scale is of several hundred years \citep{herx1_ns_free_precession_Heyl2024} which is much longer than these excursion time scales. The secular trend observed after AL3 is $-4.7(1)\times10^{-5}$ s s$^{-1}$ (Table~\ref{tab:efsearch_and_oc_analysis_table}), or roughly a $\sim0.05$\% change in period per year.

For a biaxial body ($I_3<I_2=I_1$) precessing about $I_3$, the precession frequency is given by $\omega = \Omega \Delta I_3 \cos\gamma$, where $\Delta I_3 = (I_1 - I_3)/I_3$. This is valid under the approximation that $\Omega \gg \omega$, i.e., the neutron star angular momentum is dominated by its spin, which is valid for \herx. Assuming $\Omega^{-1}\approx1.2$ s, $\Delta I_3\approx\epsilon\approx6\times10^{-7}$, and $\gamma=50^\circ$, a precession period of $\sim35$ d can be obtained \citep{kolesnikov2022evidence}. Fractional variations in $\Omega$, $\Delta I_3$, or $\gamma$ would therefore lead to comparable fractional changes in the precession period. Below, we assess these possibilities one by one.

We first consider the case in which the neutron star spin period ($P_\mathrm{spin} = \Omega^{-1} =1.2$ s) and ellipticity ($\Delta I_3=6\times10^{-7}$) remain constant. This scenario was explored by \cite{herx1_ns_free_precession_Postnov2013}, who argued that a fairly large $\gamma$ of the order of $50^\circ$ can occur from a complex non-dipolar magnetic field. In this case, the fluctuations in $\Delta I_3$ about a mean value are balanced by internal mass redistribution, while angular momentum conservation preserves the spin period. The fractional change in precession period is then given by $\frac{\Delta \suporb}{\suporb}= \mathrm{tan}\ \gamma\ \Delta \gamma$. The change in $\gamma$ then directly translates to a fractional change in precession period. For $\gamma\sim50^\circ$, a 1--10\% change in $\suporb$ would correspond to $\Delta \gamma$ between $0.48^\circ$ and $4.8^\circ$. To reproduce the observed post-AL3 secular decay, a secular decrease of just $\sim0.002^\circ$ per cycle would be required. Instantaneous changes in $\gamma$ of order $1^\circ$ have been suggested as plausible due to crustal stress variations during accretion \citep{herx1_ns_free_precession_Postnov2013}.

The second possibility is that accretion redistributes mass within the crust, modifying $\Delta I_3$. In this case, $\Delta I_3$ and precession periods vary anti-correlated, following $\frac{\delta \suporb}{\suporb}= -\frac{\delta \Delta I_3}{\Delta I_3} = -\frac{\delta \epsilon}{\epsilon}$. For an accreted mass $\delta$M onto a neutron star of mass M, $\delta \Delta I_3$ scales as $\delta M/M$. \cite{kolesnikov2022evidence} estimated that for an accretion rate of $\sim10^{17}$ g s$^{-1}$ (corresponding to $L_X\sim10^{37}$ erg s$^{-1}$), the accumulated mass over one year yields $\delta M/M\sim10^{-9}$ for a canonical 1.4 $M_\odot$ neutron star. This corresponds to a fractional change in ellipticity and hence the precession period of $\frac{\delta \epsilon}{\epsilon}=-\frac{\delta P_\mathrm{sup}}{P_\mathrm{sup}}\sim0.1$\% over a year. This value is smaller than the observed short-term stochastic variations (< 5\%) but could be comparable to, or exceed, the rate needed to account for the long-term secular decay. Therefore, accretion-induced changes in ellipticity might therefore contribute to the gradual period evolution, though likely not to the faster fluctuations.

Finally, variations in the spin period $P_\mathrm{spin}$ could, in principle, produce correlated changes in precession period, as $\frac{\Delta \suporb}{\suporb}= \frac{\Delta P_\mathrm{spin}}{P_\mathrm{spin}}$. The spin period evolution is governed by accretion torque. Long-term monitoring of \herx shows an overall long-term spin-up trend, however, at rates not exceeding $\sim-10^{-12}$ s s$^{-1}$ \footnote{\url{https://gammaray.msfc.nasa.gov/gbm/science/pulsars/lightcurves/herx1.html}} \citep{klochkov2009continuous, malacaria2020ups}. This is much smaller than the post-AL3 secular period decay. Producing the faster decay rate, stochastic fluctuations at the level of 1\% per cycle would require neutron star spin-up rates of approximately $-4\times10^{-9}$ s s$^{-1}$, at least two orders of magnitude larger than the rapid spin-up/down episodes observed in \herx \citep{malacaria2020ups}. Hence, spin-period variations alone are unlikely to dominate the precession-period changes, though minor contributions cannot be ruled out. Residual external torques over a few 100 d, as suggested by \citet{herx1_ns_free_precession_Heyl2024}, could also influence the precession period, but quantifying their contribution is beyond the scope of this work.}

\section{Summary}\label{sec:summary}
\begin{itemize}
    \item We studied the evolution of superorbital period in the well-known low mass X-ray binary pulsar, \herx, using data from five all sky monitors: \batse, \rxte, \swift, \fermi, and \maxi along with the older superorbital period data available in the literature together spanning the observation duration $\sim 50$ yr.
    
    \item Using epoch-folding search and \oc methods we found the superorbital period in \herx is decreasing at the rate of $\sim -5 \times 10^{-5}$ s s$^{-1}$. In addition to the secular decay of the superorbital period, we identified previously unreported excursions over the time scales of $\sim$10-38 superorbital cycles. These excursions indicated the presence of restoring mechanism in the system.
    
    \item We observed a positive correlation between the superorbital period and the mean count rate. This suggested that radiation-driven restoring mechanism, such as X-ray irradiation of the accretion disc or the companion star, play some role in the variations of the superorbital period.

    \item Assuming the free precession of the neutron star as the origin of the superorbital period in \herx, we find that reproducing the observed post-AL3 secular decay requires a change in the obliquity at a rate of $\sim0.002^\circ$ per cycle. However, changes in the obliquity alone cannot account for the long-term excursions observed in the superorbital period. If the observed secular decay is instead attributed to accretion-induced changes in the neutron-star ellipticity, a fractional change in ellipticity of $\sim$0.1\% is required for a mass accretion rate of $\sim10^{17}$ g s$^{-1}$. In contrast, the observed spin-up rate is insufficient to explain the secular decay of the superorbital period. Taken together, these results suggest that variations in obliquity, ellipticity, and spin period may all contribute to the observed evolution of the superorbital period.
\end{itemize}

\begin{acknowledgements}
We thank the anonymous referee for the insightful comments and suggestions, which have led to significant changes in the structure of the paper. We acknowledge the use of public data from the Swift data archive. This research has made use of MAXI data provided by \textrm{RIKEN}, \textrm{JAXA} and the \textrm{MAXI} team. This research has made use of data and/or software provided by the High Energy Astrophysics Science Archive Research Center (HEASARC), which is a service of the Astrophysics Science Division at NASA/GSFC. We acknowledge the quick-look results provided by the \textrm{ASM}/\textrm{RXTE} team. AT sincerely thanks Manish Kumar for providing the chi-square maximization code used in Sect.~\ref{sec:epoch_folding_search} and for the valuable discussions related to it.
\end{acknowledgements}

\bibliographystyle{aa}
\bibliography{references}

@ARTICLE{herx1_discovery,
       author = {{Tananbaum}, H. and {Gursky}, H. and {Kellogg}, E.~M. and {Levinson}, R. and {Schreier}, E. and {Giacconi}, R.},
        title = "{Discovery of a Periodic Pulsating Binary X-Ray Source in Hercules from UHURU}",
      journal = {\apjl},
         year = 1972,
        month = jun,
       volume = {174},
        pages = {L143},
          doi = {10.1086/180968},
       adsurl = {https://ui.adsabs.harvard.edu/abs/1972ApJ...174L.143T}
}

@article{avakyan_xrbcats_2023,
    title = {{XRBcats}: {Galactic} low-mass {X}-ray binary catalogue},
    volume = {675},
    copyright = {https://creativecommons.org/licenses/by/4.0},
    issn = {0004-6361, 1432-0746},
    shorttitle = {{XRBcats}},
    url = {https://www.aanda.org/10.1051/0004-6361/202346522},
    doi = {10.1051/0004-6361/202346522},
    urldate = {2026-04-17},
    journal = {\aap},
    author = {Avakyan, A. and Neumann, M. and Zainab, A. and Doroshenko, V. and Wilms, J. and Santangelo, A.},
    month = jul,
    year = {2023},
    pages = {A199},
}

@ARTICLE{reynolds_new_1997,
       author = {{Reynolds}, A.~P. and {Quaintrell}, H. and {Still}, M.~D. and {Roche}, P. and {Chakrabarty}, D. and {Levine}, S.~E.},
        title = "{A new mass estimate for Hercules X-1}",
      journal = {\mnras},
         year = 1997,
        month = jun,
       volume = {288},
       number = {1},
        pages = {43-52},
          doi = {10.1093/mnras/288.1.43},
       adsurl = {https://ui.adsabs.harvard.edu/abs/1997MNRAS.288...43R}
}

@ARTICLE{crampton_spectroscopic_1974,
       author = {{Crampton}, D.},
        title = "{Spectroscopic observations of HZ Herculis.}",
      journal = {\apj},
         year = 1974,
        month = jan,
       volume = {187},
        pages = {345-348},
          doi = {10.1086/152635},
       adsurl = {https://ui.adsabs.harvard.edu/abs/1974ApJ...187..345C}
}

@ARTICLE{deeter_pulse-timing_1981,
       author = {{Deeter}, J.~E. and {Boynton}, P.~E. and {Pravdo}, S.~H.},
        title = "{Pulse-timing observations of HER X-1.}",
      journal = {\apj},
         year = 1981,
        month = aug,
       volume = {247},
        pages = {1003-1012},
          doi = {10.1086/159110},
       adsurl = {https://ui.adsabs.harvard.edu/abs/1981ApJ...247.1003D}
}

@ARTICLE{middleditch_studies_1976,
       author = {{Middleditch}, J. and {Nelson}, J.},
        title = "{Studies of optical pulsations from HZ Herculis/Hercules X-1: a determination of the mass of the neutron star.}",
      journal = {\apj},
         year = 1976,
        month = sep,
       volume = {208},
        pages = {567-586},
          doi = {10.1086/154638},
       adsurl = {https://ui.adsabs.harvard.edu/abs/1976ApJ...208..567M}
}

@ARTICLE{staubert_updating_2009,
       author = {{Staubert}, R. and {Klochkov}, D. and {Wilms}, J.},
        title = "{Updating the orbital ephemeris of Hercules X-1; rate of decay and eccentricity of the orbit}",
      journal = {\aap},
         year = 2009,
        month = jun,
       volume = {500},
       number = {2},
        pages = {883-889},
          doi = {10.1051/0004-6361/200911690},
archivePrefix = {arXiv},
       eprint = {0904.2307},
 primaryClass = {astro-ph.HE},
       adsurl = {https://ui.adsabs.harvard.edu/abs/2009A&A...500..883S}
}

@ARTICLE{leahy_geometry_2025,
       author = {{Leahy}, Denis and {Mendelsohn}, Jake},
        title = "{The geometry of the Hercules X-1 accretion disk from X-rays}",
      journal = {Discover Space},
         year = 2025,
        month = mar,
       volume = {129},
       number = {1},
          eid = {5},
        pages = {5},
          doi = {10.1007/s11038-025-09564-0},
       adsurl = {https://ui.adsabs.harvard.edu/abs/2025DiSpa.129....5L}
}

@ARTICLE{katz_thirty-five-day_1973,
       author = {{Katz}, J.~I.},
        title = "{Thirty-five-day Periodicity in Her X-1}",
      journal = {Nature Physical Science},
         year = 1973,
        month = dec,
       volume = {246},
       number = {154},
        pages = {87-89},
          doi = {10.1038/physci246087a0},
       adsurl = {https://ui.adsabs.harvard.edu/abs/1973NPhS..246...87K}
}

@ARTICLE{herx1_suporb_discovery,
       author = {{Giacconi}, R. and {Gursky}, H. and {Kellogg}, E. and {Levinson}, R. and {Schreier}, E. and {Tananbaum}, H.},
        title = "{Further X-ray observations of Hercules X-1 from Uhuru.}",
      journal = {\apj},
         year = 1973,
        month = aug,
       volume = {184},
        pages = {227},
          doi = {10.1086/152321},
       adsurl = {https://ui.adsabs.harvard.edu/abs/1973ApJ...184..227G}
}

@ARTICLE{lang_read_1981,
       author = {{Lang}, F.~L. and {Levine}, A.~M. and {Bautz}, M. and {Hauskins}, S. and {Howe}, S. and {Primini}, F.~A. and {Lewin}, W.~H.~G. and {Baity}, W.~A. and {Knight}, F.~K. and {Rotschild}, R.~E. and {Petterson}, J.~A.},
        title = "{Discovery of a 30.5 day periodicity in LMC X-4.}",
      journal = {\apjl},
         year = 1981,
        month = may,
       volume = {246},
        pages = {L21-L25},
          doi = {10.1086/183545},
       adsurl = {https://ui.adsabs.harvard.edu/abs/1981ApJ...246L..21L}
}

@article{wojdowski_quasiperiodic_1998,
    title = {Quasi‐periodic {Occultation} by a {Precessing} {Accretion} {Disk} and {Other} {Variabilities} of {SMC} {X}‐1},
    volume = {502},
    issn = {0004-637X, 1538-4357},
    url = {https://iopscience.iop.org/article/10.1086/305893},
    doi = {10.1086/305893},
    language = {en},
    number = {1},
    urldate = {2025-06-27},
    journal = {\apj},
    author = {Wojdowski, Patrick and Clark, George W. and Levine, Alan M. and Woo, Jonathan W. and Zhang, Shuang Nan},
    month = jul,
    year = {1998},
    pages = {253--264},
}

@ARTICLE{trowbridge2007tracking,
       author = {{Trowbridge}, Sarah and {Nowak}, Michael A. and {Wilms}, J{\"o}rn},
        title = "{Tracking the Orbital and Superorbital Periods of SMC X-1}",
      journal = {\apj},
         year = 2007,
        month = nov,
       volume = {670},
       number = {1},
        pages = {624-634},
          doi = {10.1086/522075},
archivePrefix = {arXiv},
       eprint = {0708.0038},
 primaryClass = {astro-ph},
       adsurl = {https://ui.adsabs.harvard.edu/abs/2007ApJ...670..624T}
}

@ARTICLE{hjellming_analysis_1981,
       author = {{Hjellming}, R.~M. and {Johnston}, K.~J.},
        title = "{An analysis of the proper motions of SS 433 radio jets.}",
      journal = {\apjl},
         year = 1981,
        month = jun,
       volume = {246},
        pages = {L141-L145},
          doi = {10.1086/183571},
       adsurl = {https://ui.adsabs.harvard.edu/abs/1981ApJ...246L.141H}
}

@ARTICLE{margon_observations_1984,
       author = {{Margon}, Bruce},
        title = "{Observations of SS 433}",
      journal = {\araa},
         year = 1984,
        month = jan,
       volume = {22},
        pages = {507-536},
          doi = {10.1146/annurev.aa.22.090184.002451},
       adsurl = {https://ui.adsabs.harvard.edu/abs/1984ARA&A..22..507M}
}

@ARTICLE{scott_rxte_asm_35d,
       author = {{Leahy}, D. Matthew Scott Denis A.},
        title = "{RXTE/ASM Observations of the 35-day Cycle of Her X-1}",
      journal = {arXiv e-prints},
         year = 1998,
        month = sep,
          eid = {astro-ph/9809152},
        pages = {astro-ph/9809152},
          doi = {10.48550/arXiv.astro-ph/9809152},
archivePrefix = {arXiv},
       eprint = {astro-ph/9809152},
 primaryClass = {astro-ph},
       adsurl = {https://ui.adsabs.harvard.edu/abs/1998astro.ph..9152L}
}

@ARTICLE{leahy_ginga_2000,
       author = {{Leahy}, D.~A.},
        title = "{Ginga observations of Her X-1 in the short high state: evidence for an extended emission region}",
      journal = {\mnras},
         year = 2000,
        month = jul,
       volume = {315},
       number = {4},
        pages = {735-739},
          doi = {10.1046/j.1365-8711.2000.03460.x},
       adsurl = {https://ui.adsabs.harvard.edu/abs/2000MNRAS.315..735L}
}

@ARTICLE{staubert_inversion_2017,
       author = {{Staubert}, R. and {Klochkov}, D. and {F{\"u}rst}, F. and {Wilms}, J. and {Rothschild}, R.~E. and {Harrison}, F.},
        title = "{Inversion of the decay of the cyclotron line energy in Hercules X-1}",
      journal = {\aap},
         year = 2017,
        month = oct,
       volume = {606},
          eid = {L13},
        pages = {L13},
          doi = {10.1051/0004-6361/201731927},
archivePrefix = {arXiv},
       eprint = {1710.05150},
 primaryClass = {astro-ph.HE},
       adsurl = {https://ui.adsabs.harvard.edu/abs/2017A&A...606L..13S}
}

@ARTICLE{gerend_optical_1976,
       author = {{Gerend}, D. and {Boynton}, P.~E.},
        title = "{Optical clues to the nature of Hercules X-1 / HZ Herculis.}",
      journal = {\apj},
         year = 1976,
        month = oct,
       volume = {209},
        pages = {562-573},
          doi = {10.1086/154751},
       adsurl = {https://ui.adsabs.harvard.edu/abs/1976ApJ...209..562G}
}

@ARTICLE{scott_35_2000,
       author = {{Scott}, D. Matthew and {Leahy}, Denis A. and {Wilson}, Robert B.},
        title = "{The 35 Day Evolution of the Hercules X-1 Pulse Profile: Evidence for a Resolved Inner Disk Occultation of the Neutron Star}",
      journal = {\apj},
         year = 2000,
        month = aug,
       volume = {539},
       number = {1},
        pages = {392-412},
          doi = {10.1086/309203},
archivePrefix = {arXiv},
       eprint = {astro-ph/0002327},
 primaryClass = {astro-ph},
       adsurl = {https://ui.adsabs.harvard.edu/abs/2000ApJ...539..392S}
}

@ARTICLE{leahy_modelling_2002,
       author = {{Leahy}, D.~A.},
        title = "{Modelling RXTE/ASM observations of the 35-d cycle in Her X-1}",
      journal = {\mnras},
         year = 2002,
        month = aug,
       volume = {334},
       number = {4},
        pages = {847-854},
          doi = {10.1046/j.1365-8711.2002.05547.x},
       adsurl = {https://ui.adsabs.harvard.edu/abs/2002MNRAS.334..847L}
}

@ARTICLE{leahy_35-day_2021,
       author = {{Leahy}, Denis and {Wang}, Yuyang},
        title = "{The 35-Day Cycle of Hercules X-1 in Multiple Energy Bands from MAXI and Swift/BAT Monitoring}",
      journal = {Universe},
         year = 2021,
        month = may,
       volume = {7},
       number = {6},
          eid = {160},
        pages = {160},
          doi = {10.3390/universe7060160},
       adsurl = {https://ui.adsabs.harvard.edu/abs/2021Univ....7..160L}
}

@article{brecher_precession_1975,
    title = {Precession in transient {X}-ray sources},
    volume = {257},
    copyright = {https://www.springernature.com/gp/researchers/text-and-data-mining},
    issn = {0028-0836, 1476-4687},
    url = {https://www.nature.com/articles/257203a0},
    doi = {10.1038/257203a0},
    language = {en},
    number = {5523},
    urldate = {2025-06-08},
    journal = {\nat},
    author = {Brecher, Kenneth},
    month = sep,
    year = {1975},
    pages = {203--203},
}

@ARTICLE{herx1_ns_free_precession_Truemper1986,
       author = {{Truemper}, J. and {Kahabka}, P. and {Oegelman}, H. and {Pietsch}, W. and {Voges}, W.},
        title = "{EXOSAT Observations of the 35 Day Cycle of Hercules X-1: Evidence for Neutron Star Precession}",
      journal = {\apjl},
         year = 1986,
        month = jan,
       volume = {300},
        pages = {L63},
          doi = {10.1086/184604},
       adsurl = {https://ui.adsabs.harvard.edu/abs/1986ApJ...300L..63T}
}

@ARTICLE{klochkov_integral_2008,
       author = {{Klochkov}, D. and {Staubert}, R. and {Postnov}, K. and {Shakura}, N. and {Santangelo}, A. and {Tsygankov}, S. and {Lutovinov}, A. and {Kreykenbohm}, I. and {Wilms}, J.},
        title = "{INTEGRAL observations of Hercules X-1}",
      journal = {\aap},
         year = 2008,
        month = may,
       volume = {482},
       number = {3},
        pages = {907-915},
          doi = {10.1051/0004-6361:20078953},
archivePrefix = {arXiv},
       eprint = {0802.1168},
 primaryClass = {astro-ph},
       adsurl = {https://ui.adsabs.harvard.edu/abs/2008A&A...482..907K}
}

@ARTICLE{herx1_ns_free_precession_Postnov2013,
       author = {{Postnov}, K. and {Shakura}, N. and {Staubert}, R. and {Kochetkova}, A. and {Klochkov}, D. and {Wilms}, J.},
        title = "{Variable neutron star free precession in Hercules X-1 from evolution of RXTE X-ray pulse profiles with phase of the 35-d cycle}",
      journal = {\mnras},
         year = 2013,
        month = oct,
       volume = {435},
       number = {2},
        pages = {1147-1164},
          doi = {10.1093/mnras/stt1363},
archivePrefix = {arXiv},
       eprint = {1307.6026},
 primaryClass = {astro-ph.HE},
       adsurl = {https://ui.adsabs.harvard.edu/abs/2013MNRAS.435.1147P}
}

@ARTICLE{herx1_ns_free_precession_Heyl2024,
       author = {{Heyl}, Jeremy and {Doroshenko}, Victor and {Gonz{\'a}lez-Caniulef}, Denis and {Caiazzo}, Ilaria and {Poutanen}, Juri and {Mushtukov}, Alexander and {Tsygankov}, Sergey S. and {Kirmizibayrak}, Demet and {Bachetti}, Matteo and {Pavlov}, George G. and {Forsblom}, Sofia V. and {Malacaria}, Christian and {Suleimanov}, Valery F. and {Agudo}, Iv{\'a}n and {Antonelli}, Lucio Angelo and {Baldini}, Luca and {Baumgartner}, Wayne H. and {Bellazzini}, Ronaldo and {Bianchi}, Stefano and {Bongiorno}, Stephen D. and {Bonino}, Raffaella and {Brez}, Alessandro and {Bucciantini}, Niccol{\`o} and {Capitanio}, Fiamma and {Castellano}, Simone and {Cavazzuti}, Elisabetta and {Chen}, Chien-Ting and {Ciprini}, Stefano and {Costa}, Enrico and {De Rosa}, Alessandra and {Del Monte}, Ettore and {Di Gesu}, Laura and {Di Lalla}, Niccol{\`o} and {Di Marco}, Alessandro and {Donnarumma}, Immacolata and {Dov{\v{c}}iak}, Michal and {Ehlert}, Steven R. and {Enoto}, Teruaki and {Evangelista}, Yuri and {Fabiani}, Sergio and {Ferrazzoli}, Riccardo and {Garcia}, Javier A. and {Gunji}, Shuichi and {Hayashida}, Kiyoshi and {Iwakiri}, Wataru and {Jorstad}, Svetlana G. and {Kaaret}, Philip and {Karas}, Vladimir and {Kislat}, Fabian and {Kitaguchi}, Takao and {Kolodziejczak}, Jeffery J. and {Krawczynski}, Henric and {La Monaca}, Fabio and {Latronico}, Luca and {Liodakis}, Ioannis and {Maldera}, Simone and {Manfreda}, Alberto and {Marin}, Fr{\'e}d{\'e}ric and {Marinucci}, Andrea and {Marscher}, Alan P. and {Marshall}, Herman L. and {Massaro}, Francesco and {Matt}, Giorgio and {Mitsuishi}, Ikuyuki and {Mizuno}, Tsunefumi and {Muleri}, Fabio and {Negro}, Michela and {Ng}, C.-Y. and {O'Dell}, Stephen L. and {Omodei}, Nicola and {Oppedisano}, Chiara and {Papitto}, Alessandro and {Peirson}, Abel Lawrence and {Perri}, Matteo and {Pesce-Rollins}, Melissa and {Petrucci}, Pierre-Olivier and {Pilia}, Maura and {Possenti}, Andrea and {Puccetti}, Simonetta and {Ramsey}, Brian D. and {Rankin}, John and {Ratheesh}, Ajay and {Roberts}, Oliver J. and {Romani}, Roger W. and {Sgr{\`o}}, Carmelo and {Slane}, Patrick and {Soffitta}, Paolo and {Spandre}, Gloria and {Swartz}, Douglas A. and {Tamagawa}, Toru and {Tavecchio}, Fabrizio and {Taverna}, Roberto and {Tawara}, Yuzuru and {Tennant}, Allyn F. and {Thomas}, Nicholas E. and {Tombesi}, Francesco and {Trois}, Alessio and {Turolla}, Roberto and {Vink}, Jacco and {Weisskopf}, Martin C. and {Wu}, Kinwah and {Xie}, Fei and {Zane}, Silvia},
        title = "{Complex rotational dynamics of the neutron star in Hercules X-1 revealed by X-ray polarization}",
      journal = {Nat. Astron.},
         year = 2024,
        month = aug,
       volume = {8},
        pages = {1047-1053},
          doi = {10.1038/s41550-024-02295-8},
       adsurl = {https://ui.adsabs.harvard.edu/abs/2024NatAs...8.1047H}
}

@ARTICLE{STA83,
       author = {{Staubert}, R. and {Bezler}, M. and {Kendziorra}, E.},
        title = "{Hercules X-1 : a random walk noise model for the 35-day turn-ons.}",
      journal = {\aap},
         year = 1983,
        month = jan,
       volume = {117},
        pages = {215-219},
       adsurl = {https://ui.adsabs.harvard.edu/abs/1983A&A...117..215S}
}

@ARTICLE{baykal_1993,
       author = {{Baykal}, A. and {Boynton}, P.~E. and {Deeter}, J.~E. and {Scott}, D.~M.},
        title = "{The noise in the 35-d cycle of HER X-1.}",
      journal = {\mnras},
         year = 1993,
        month = nov,
       volume = {265},
        pages = {347-350},
          doi = {10.1093/mnras/265.2.347},
       adsurl = {https://ui.adsabs.harvard.edu/abs/1993MNRAS.265..347B}
}

@ARTICLE{leahy_swiftbat_2020,
       author = {{Leahy}, Denis and {Wang}, Yuyang},
        title = "{Swift/BAT and RXTE/ASM Observations of the 35 day X-Ray Cycle of Hercules X-1}",
      journal = {\apj},
         year = 2020,
        month = oct,
       volume = {902},
       number = {2},
          eid = {146},
        pages = {146},
          doi = {10.3847/1538-4357/abb611},
archivePrefix = {arXiv},
       eprint = {2009.07246},
 primaryClass = {astro-ph.HE},
       adsurl = {https://ui.adsabs.harvard.edu/abs/2020ApJ...902..146L}
}

@INPROCEEDINGS{STA06,
       author = {{Staubert}, R. and {Schandl}, S. and {Klochkov}, D. and {Wilms}, J. and {Postnov}, K. and {Shakura}, N.},
        title = "{Long-term developments in Her X-1: Correlation between the histories of the 35 day turn-on cycle and the 1.24 sec pulse period}",
    booktitle = {The Transient Milky Way: A Perspective for MIRAX},
         year = 2006,
       editor = {{D'Amico}, Flavio and {Braga}, Jo{\~a}o and {Rothschild}, Richard E.},
       series = {American Institute of Physics Conference Series},
       volume = {840},
        month = jun,
    publisher = {AIP},
        pages = {65-70},
          doi = {10.1063/1.2216605},
archivePrefix = {arXiv},
       eprint = {astro-ph/0702528},
 primaryClass = {astro-ph},
       adsurl = {https://ui.adsabs.harvard.edu/abs/2006AIPC..840...65S}
}

@ARTICLE{batse_meegan,
       author = {{Meegan}, C.~A. and {Fishman}, G.~J. and {Wilson}, R.~B. and {Paciesas}, W.~S. and {Pendleton}, G.~N. and {Horack}, J.~M. and {Brock}, M.~N. and {Kouveliotou}, C.},
        title = "{Spatial distribution of {\ensuremath{\gamma}}-ray bursts observed by BATSE}",
      journal = {\nat},
         year = 1992,
        month = jan,
       volume = {355},
       number = {6356},
        pages = {143-145},
          doi = {10.1038/355143a0},
       adsurl = {https://ui.adsabs.harvard.edu/abs/1992Natur.355..143M}
}

@ARTICLE{asm_levine,
       author = {{Levine}, Alan M. and {Bradt}, Hale and {Cui}, Wei and {Jernigan}, J.~G. and {Morgan}, Edward H. and {Remillard}, Ronald and {Shirey}, Robert E. and {Smith}, Donald A.},
        title = "{First Results from the All-Sky Monitor on the Rossi X-Ray Timing Explorer}",
      journal = {\apjl},
         year = 1996,
        month = sep,
       volume = {469},
        pages = {L33},
          doi = {10.1086/310260},
archivePrefix = {arXiv},
       eprint = {astro-ph/9608109},
 primaryClass = {astro-ph},
       adsurl = {https://ui.adsabs.harvard.edu/abs/1996ApJ...469L..33L}
}

@INPROCEEDINGS{rxte_jahoda,
       author = {{Jahoda}, Keith and {Swank}, Jean H. and {Giles}, Alan B. and {Stark}, Michael J. and {Strohmayer}, Tod and {Zhang}, Weiping and {Morgan}, Edward H.},
        title = "{In-orbit performance and calibration of the Rossi X-ray Timing Explorer (RXTE) Proportional Counter Array (PCA)}",
    booktitle = {EUV, X-Ray, and Gamma-Ray Instrumentation for Astronomy VII},
         year = 1996,
       editor = {{Siegmund}, Oswald H. and {Gummin}, Mark A.},
       series = {Society of Photo-Optical Instrumentation Engineers (SPIE) Conference Series},
       volume = {2808},
        month = oct,
        pages = {59-70},
          doi = {10.1117/12.256034},
       adsurl = {https://ui.adsabs.harvard.edu/abs/1996SPIE.2808...59J}
}

@ARTICLE{Swift_Gehrels_2004,
       author = {{Gehrels}, N. and {Chincarini}, G. and {Giommi}, P. and {Mason}, K.~O. and {Nousek}, J.~A. and {Wells}, A.~A. and {White}, N.~E. and {Barthelmy}, S.~D. and {Burrows}, D.~N. and {Cominsky}, L.~R. and {Hurley}, K.~C. and {Marshall}, F.~E. and {M{\'e}sz{\'a}ros}, P. and {Roming}, P.~W.~A. and {Angelini}, L. and {Barbier}, L.~M. and {Belloni}, T. and {Campana}, S. and {Caraveo}, P.~A. and {Chester}, M.~M. and {Citterio}, O. and {Cline}, T.~L. and {Cropper}, M.~S. and {Cummings}, J.~R. and {Dean}, A.~J. and {Feigelson}, E.~D. and {Fenimore}, E.~E. and {Frail}, D.~A. and {Fruchter}, A.~S. and {Garmire}, G.~P. and {Gendreau}, K. and {Ghisellini}, G. and {Greiner}, J. and {Hill}, J.~E. and {Hunsberger}, S.~D. and {Krimm}, H.~A. and {Kulkarni}, S.~R. and {Kumar}, P. and {Lebrun}, F. and {Lloyd-Ronning}, N.~M. and {Markwardt}, C.~B. and {Mattson}, B.~J. and {Mushotzky}, R.~F. and {Norris}, J.~P. and {Osborne}, J. and {Paczynski}, B. and {Palmer}, D.~M. and {Park}, H.-S. and {Parsons}, A.~M. and {Paul}, J. and {Rees}, M.~J. and {Reynolds}, C.~S. and {Rhoads}, J.~E. and {Sasseen}, T.~P. and {Schaefer}, B.~E. and {Short}, A.~T. and {Smale}, A.~P. and {Smith}, I.~A. and {Stella}, L. and {Tagliaferri}, G. and {Takahashi}, T. and {Tashiro}, M. and {Townsley}, L.~K. and {Tueller}, J. and {Turner}, M.~J.~L. and {Vietri}, M. and {Voges}, W. and {Ward}, M.~J. and {Willingale}, R. and {Zerbi}, F.~M. and {Zhang}, W.~W.},
        title = "{The Swift Gamma-Ray Burst Mission}",
      journal = {\apj},
         year = 2004,
        month = aug,
       volume = {611},
       number = {2},
        pages = {1005-1020},
          doi = {10.1086/422091},
archivePrefix = {arXiv},
       eprint = {astro-ph/0405233},
 primaryClass = {astro-ph},
       adsurl = {https://ui.adsabs.harvard.edu/abs/2004ApJ...611.1005G}
}

@ARTICLE{bat_2005,
       author = {{Barthelmy}, Scott D. and {Barbier}, Louis M. and {Cummings}, Jay R. and {Fenimore}, Ed E. and {Gehrels}, Neil and {Hullinger}, Derek and {Krimm}, Hans A. and {Markwardt}, Craig B. and {Palmer}, David M. and {Parsons}, Ann and {Sato}, Goro and {Suzuki}, Masaya and {Takahashi}, Tadayuki and {Tashiro}, Makota and {Tueller}, Jack},
        title = "{The Burst Alert Telescope (BAT) on the SWIFT Midex Mission}",
      journal = {\ssr},
         year = 2005,
        month = oct,
       volume = {120},
       number = {3-4},
        pages = {143-164},
          doi = {10.1007/s11214-005-5096-3},
archivePrefix = {arXiv},
       eprint = {astro-ph/0507410},
 primaryClass = {astro-ph},
       adsurl = {https://ui.adsabs.harvard.edu/abs/2005SSRv..120..143B}
}

@ARTICLE{meegan2009fermi,
       author = {{Meegan}, Charles and {Lichti}, Giselher and {Bhat}, P.~N. and {Bissaldi}, Elisabetta and {Briggs}, Michael S. and {Connaughton}, Valerie and {Diehl}, Roland and {Fishman}, Gerald and {Greiner}, Jochen and {Hoover}, Andrew S. and {van der Horst}, Alexander J. and {von Kienlin}, Andreas and {Kippen}, R. Marc and {Kouveliotou}, Chryssa and {McBreen}, Sheila and {Paciesas}, W.~S. and {Preece}, Robert and {Steinle}, Helmut and {Wallace}, Mark S. and {Wilson}, Robert B. and {Wilson-Hodge}, Colleen},
        title = "{The Fermi Gamma-ray Burst Monitor}",
      journal = {\apj},
         year = 2009,
        month = sep,
       volume = {702},
       number = {1},
        pages = {791-804},
          doi = {10.1088/0004-637X/702/1/791},
archivePrefix = {arXiv},
       eprint = {0908.0450},
 primaryClass = {astro-ph.IM},
       adsurl = {https://ui.adsabs.harvard.edu/abs/2009ApJ...702..791M}
}

@ARTICLE{maxi,
       author = {{Matsuoka}, Masaru and {Kawasaki}, Kazuyoshi and {Ueno}, Shiro and {Tomida}, Hiroshi and {Kohama}, Mitsuhiro and {Suzuki}, Motoko and {Adachi}, Yasuki and {Ishikawa}, Masaki and {Mihara}, Tatehiro and {Sugizaki}, Mutsumi and {Isobe}, Naoki and {Nakagawa}, Yujin and {Tsunemi}, Hiroshi and {Miyata}, Emi and {Kawai}, Nobuyuki and {Kataoka}, Jun and {Morii}, Mikio and {Yoshida}, Atsumasa and {Negoro}, Hitoshi and {Nakajima}, Motoki and {Ueda}, Yoshihiro and {Chujo}, Hirotaka and {Yamaoka}, Kazutaka and {Yamazaki}, Osamu and {Nakahira}, Satoshi and {You}, Tetsuya and {Ishiwata}, Ryoji and {Miyoshi}, Sho and {Eguchi}, Satoshi and {Hiroi}, Kazuo and {Katayama}, Haruyoshi and {Ebisawa}, Ken},
        title = "{The MAXI Mission on the ISS: Science and Instruments for Monitoring All-Sky X-Ray Images}",
      journal = {\pasj},
         year = 2009,
        month = oct,
       volume = {61},
        pages = {999},
          doi = {10.1093/pasj/61.5.999},
archivePrefix = {arXiv},
       eprint = {0906.0631},
 primaryClass = {astro-ph.IM},
       adsurl = {https://ui.adsabs.harvard.edu/abs/2009PASJ...61..999M}
}

@ARTICLE{gsc_maxi,
       author = {{Mihara}, Tatehiro and {Nakajima}, Motoki and {Sugizaki}, Mutsumi and {Serino}, Motoko and {Matsuoka}, Masaru and {Kohama}, Mitsuhiro and {Kawasaki}, Kazuyoshi and {Tomida}, Hiroshi and {Ueno}, Shiro and {Kawai}, Nobuyuki and {Kataoka}, Jun and {Morii}, Mikio and {Yoshida}, Atsumasa and {Yamaoka}, Kazutaka and {Nakahira}, Satoshi and {Negoro}, Hitoshi and {Isobe}, Naoki and {Yamauchi}, Makoto and {Sakurai}, Ikuya},
        title = "{Gas Slit Camera (GSC) onboard MAXI on ISS}",
      journal = {\pasj},
         year = 2011,
        month = nov,
       volume = {63},
        pages = {S623-S634},
          doi = {10.1093/pasj/63.sp3.S623},
archivePrefix = {arXiv},
       eprint = {1103.4224},
 primaryClass = {astro-ph.IM},
       adsurl = {https://ui.adsabs.harvard.edu/abs/2011PASJ...63S.623M}
}

@incollection{di_gesu_xronos_1992,
    address = {Boston, MA},
    title = {{XRONOS}: {A} {Timing} {Analysis} {Software} {Package}},
    isbn = {978-1-4613-6496-2 978-1-4615-3388-7},
    shorttitle = {{XRONOS}},
    url = {http://link.springer.com/10.1007/978-1-4615-3388-7_7},
    doi = {10.1007/978-1-4615-3388-7_7},
    language = {en},
    urldate = {2026-05-29},
    booktitle = {Data {Analysis} in {Astronomy} {IV}},
    publisher = {Springer US},
    author = {Stella, L. and Angelini, L.},
    editor = {Di Gesù, V. and Scarsi, L. and Buccheri, R. and Crane, P. and Maccarone, M. C. and Zimmermann, H. U.},
    year = {1992},
    pages = {59--64},
}

@ARTICLE{clarkson_I,
       author = {{Clarkson}, W.~I. and {Charles}, P.~A. and {Coe}, M.~J. and {Laycock}, S. and {Tout}, M.~D. and {Wilson}, C.~A.},
        title = "{Long-term properties of accretion discs in X-ray binaries - I. The variable third period in SMC X-1}",
      journal = {\mnras},
         year = 2003,
        month = feb,
       volume = {339},
       number = {2},
        pages = {447-454},
          doi = {10.1046/j.1365-8711.2003.06176.x},
archivePrefix = {arXiv},
       eprint = {astro-ph/0210241},
 primaryClass = {astro-ph},
       adsurl = {https://ui.adsabs.harvard.edu/abs/2003MNRAS.339..447C}
}

@ARTICLE{hu_superorbital_2013,
       author = {{Hu}, Chin-Ping and {Chou}, Yi and {Yang}, Ting-Chang and {Su}, Yi-Hao},
        title = "{Superorbital Phase-resolved Analysis of SMC X-1}",
      journal = {\apj},
         year = 2013,
        month = aug,
       volume = {773},
       number = {1},
          eid = {58},
        pages = {58},
          doi = {10.1088/0004-637X/773/1/58},
archivePrefix = {arXiv},
       eprint = {1306.5819},
 primaryClass = {astro-ph.HE},
       adsurl = {https://ui.adsabs.harvard.edu/abs/2013ApJ...773...58H}
}

@ARTICLE{singh_cygx1_2002,
       author = {{Singh}, N.~S. and {Naik}, S. and {Paul}, B. and {Agrawal}, P.~C. and {Rao}, A.~R. and {Singh}, K.~Y.},
        title = "{New measurements of orbital period change in Cygnus X-3}",
      journal = {\aap},
         year = 2002,
        month = sep,
       volume = {392},
        pages = {161-167},
          doi = {10.1051/0004-6361:20020923},
archivePrefix = {arXiv},
       eprint = {astro-ph/0206134},
 primaryClass = {astro-ph},
       adsurl = {https://ui.adsabs.harvard.edu/abs/2002A&A...392..161S}
}

@ARTICLE{paul_superorbital_2002,
       author = {{Paul}, B. and {Kitamoto}, S.},
        title = "{Superorbital Period Variations in the X-ray Pulsar LMC X-4}",
      journal = {JA\&A},
         year = 2002,
        month = mar,
       volume = {23},
       number = {1-2},
        pages = {33-38},
          doi = {10.1007/BF02702462},
       adsurl = {https://ui.adsabs.harvard.edu/abs/2002JApA...23...33P}
}

@ARTICLE{lmcx4_orbPdot_Naik_2004,
       author = {{Naik}, S. and {Paul}, B.},
        title = "{Timing and Spectral Studies of LMC X-4 in High and Low States with BeppoSAX: Detection of Pulsations in the Soft Spectral Component}",
      journal = {\apj},
         year = 2004,
        month = jan,
       volume = {600},
       number = {1},
        pages = {351-357},
          doi = {10.1086/379803},
archivePrefix = {arXiv},
       eprint = {astro-ph/0309431},
 primaryClass = {astro-ph},
       adsurl = {https://ui.adsabs.harvard.edu/abs/2004ApJ...600..351N}
}

@ARTICLE{cenx3_smcx1_orbPdot_raichur_2010,
       author = {{Raichur}, Harsha and {Paul}, Biswajit},
        title = "{Effect of pulse profile variations on measurement of eccentricity in orbits of Cen X-3 and SMC X-1}",
      journal = {\mnras},
         year = 2010,
        month = jan,
       volume = {401},
       number = {3},
        pages = {1532-1539},
          doi = {10.1111/j.1365-2966.2009.15778.x},
archivePrefix = {arXiv},
       eprint = {0909.4271},
 primaryClass = {astro-ph.HE},
       adsurl = {https://ui.adsabs.harvard.edu/abs/2010MNRAS.401.1532R}
}

@ARTICLE{manikantan_gx301m2,
       author = {{Manikantan}, Hemanth and {Kumar}, Manish and {Paul}, Biswajit and {Rana}, Vikram},
        title = "{Investigating the orbital evolution of the eccentric HMXB GX 301-2 using long-term X-ray light curves}",
      journal = {\mnras},
         year = 2024,
        month = jan,
       volume = {527},
       number = {1},
        pages = {640-650},
          doi = {10.1093/mnras/stad3090},
archivePrefix = {arXiv},
       eprint = {2310.02626},
 primaryClass = {astro-ph.HE},
       adsurl = {https://ui.adsabs.harvard.edu/abs/2024MNRAS.527..640M}
}

@article{ramer_iterative_1972,
    title = {An iterative procedure for the polygonal approximation of plane curves},
    volume = {1},
    copyright = {https://www.elsevier.com/tdm/userlicense/1.0/},
    issn = {0146664X},
    url = {https://linkinghub.elsevier.com/retrieve/pii/S0146664X72800170},
    doi = {10.1016/S0146-664X(72)80017-0},
    language = {en},
    number = {3},
    urldate = {2026-09-09},
    journal = {Computer Graphics and Image Processing},
    author = {Ramer, Urs},
    month = nov,
    year = {1972},
    pages = {244--256},
}

@ARTICLE{douglas_algorithms_1973,
       author = {{Douglas}, David H. and {Peucker}, Thomas K.},
        title = "{Algorithms for the Reduction of the Number of Points Required to Represent a Digitized Line or its Caricature}",
      journal = {Cartographica: The International Journal for Geographic Information and Geovisualization},
         year = 1973,
        month = dec,
       volume = {10},
       number = {2},
        pages = {112-122},
          doi = {10.3138/FM57-6770-U75U-7727},
       adsurl = {https://ui.adsabs.harvard.edu/abs/1973CIJGI..10..112D}
}

@ARTICLE{jain_eclipse_2022,
       author = {{Jain}, Chetana and {Sharma}, Rahul and {Paul}, Biswajit},
        title = "{Eclipse Timings of the LMXB XTE J1710-281 : discovery of a third orbital period glitch}",
      journal = {\mnras},
         year = 2022,
        month = dec,
       volume = {517},
       number = {2},
        pages = {2131-2137},
          doi = {10.1093/mnras/stac2804},
archivePrefix = {arXiv},
       eprint = {2209.13406},
 primaryClass = {astro-ph.HE},
       adsurl = {https://ui.adsabs.harvard.edu/abs/2022MNRAS.517.2131J}
}

@ARTICLE{liao_torque_2022,
       author = {{Liao}, Zhenxuan and {Liu}, Jiren and {Gou}, Lijun},
        title = "{Torque reversals and wind variations of X-ray pulsar Vela X-1}",
      journal = {\mnras},
         year = 2022,
        month = nov,
       volume = {517},
       number = {1},
        pages = {L111-L115},
          doi = {10.1093/mnrasl/slac119},
archivePrefix = {arXiv},
       eprint = {2210.02032},
 primaryClass = {astro-ph.HE},
       adsurl = {https://ui.adsabs.harvard.edu/abs/2022MNRAS.517L.111L}
}

@ARTICLE{liao_torque-dependent_2023,
       author = {{Liao}, Zhenxuan and {Liu}, Jiren},
        title = "{Torque-dependent orbital modulation of X-ray pulsar Cen X-3}",
      journal = {\mnras},
         year = 2024,
        month = mar,
       volume = {529},
       number = {1},
        pages = {L130-L134},
          doi = {10.1093/mnrasl/slae009},
archivePrefix = {arXiv},
       eprint = {2401.09197},
 primaryClass = {astro-ph.HE},
       adsurl = {https://ui.adsabs.harvard.edu/abs/2024MNRAS.529L.130L}
}

@ARTICLE{delgado_al1_hz_1983,
       author = {{Delgado}, A.~J. and {Schmidt}, H.~U. and {Thomas}, H.-C.},
        title = "{HZ Herculis, still active.}",
      journal = {\aap},
         year = 1983,
        month = nov,
       volume = {127},
        pages = {L15-L16},
       adsurl = {https://ui.adsabs.harvard.edu/abs/1983A&A...127L..15D}
}

@ARTICLE{parmar_al1_an_1985,
       author = {{Parmar}, A.~N. and {Pietsch}, W. and {McKechnie}, S. and {White}, N.~E. and {Truemper}, J. and {Voges}, W. and {Barr}, P.},
        title = "{An extended X-ray low state from Hercules X-1}",
      journal = {\nat},
         year = 1985,
        month = jan,
       volume = {313},
       number = {5998},
        pages = {119-121},
          doi = {10.1038/313119a0},
       adsurl = {https://ui.adsabs.harvard.edu/abs/1985Natur.313..119P}
}

@ARTICLE{vrtilek_al2_multiwavelength_1994,
       author = {{Vrtilek}, S.~D. and {Mihara}, T. and {Primini}, F.~A. and {Kahabka}, P. and {Marshall}, H. and {Agerer}, F. and {Charles}, P.~A. and {Cheng}, F.~H. and {Dennerl}, K. and {La Dous}, C. and {Hu}, E.~M. and {Rutten}, R. and {Serlemitsos}, P. and {Soong}, Y. and {Stull}, J. and {Truemper}, J. and {Voges}, W. and {Wagner}, R.~M. and {Wilson}, R.},
        title = "{Multiwavelength Observations of Hercules X-1/HZ Herculis}",
      journal = {\apjl},
         year = 1994,
        month = nov,
       volume = {436},
        pages = {L9},
          doi = {10.1086/187621},
       adsurl = {https://ui.adsabs.harvard.edu/abs/1994ApJ...436L...9V}
}

@ARTICLE{parmar_al3_x-ray_1999,
       author = {{Parmar}, A.~N. and {Oosterbroek}, T. and {dal Fiume}, D. and {Orlandini}, M. and {Santangelo}, A. and {Segreto}, A. and {del Sordo}, S.},
        title = "{X-ray observations during a HER X-1 anomalous low-state}",
      journal = {\aap},
         year = 1999,
        month = oct,
       volume = {350},
        pages = {L5-L8},
          doi = {10.48550/arXiv.astro-ph/9909039},
archivePrefix = {arXiv},
       eprint = {astro-ph/9909039},
 primaryClass = {astro-ph},
       adsurl = {https://ui.adsabs.harvard.edu/abs/1999A&A...350L...5P}
}

@ARTICLE{coburn_al3_the_2000,
       author = {{Coburn}, W. and {Heindl}, W.~A. and {Wilms}, J. and {Gruber}, D.~E. and {Staubert}, R. and {Rothschild}, R.~E. and {Postnov}, K.~A. and {Shakura}, N. and {Risse}, P. and {Kreykenbohm}, I. and {Pelling}, M.~R.},
        title = "{The 1999 Hercules X-1 Anomalous Low State}",
      journal = {\apj},
         year = 2000,
        month = nov,
       volume = {543},
       number = {1},
        pages = {351-358},
          doi = {10.1086/317081},
archivePrefix = {arXiv},
       eprint = {astro-ph/9911533},
 primaryClass = {astro-ph},
       adsurl = {https://ui.adsabs.harvard.edu/abs/2000ApJ...543..351C}
}

@ARTICLE{still_al3_rxte_2001,
       author = {{Still}, Martin and {O'Brien}, Kieran and {Horne}, Keith and {Hudson}, Danny and {Boroson}, Bram and {Vrtilek}, Saeqa Dil and {Quaintrell}, Hannah and {Fiedler}, Hauke},
        title = "{RXTE Observations of Hercules X-1 during the 1998 July Short High State}",
      journal = {\apj},
         year = 2001,
        month = jun,
       volume = {553},
       number = {2},
        pages = {776-787},
          doi = {10.1086/320951},
archivePrefix = {arXiv},
       eprint = {astro-ph/0102100},
 primaryClass = {astro-ph},
       adsurl = {https://ui.adsabs.harvard.edu/abs/2001ApJ...553..776S}
}

@ARTICLE{staubert_two_2009,
       author = {{Staubert}, R. and {Klochkov}, D. and {Postnov}, K. and {Shakura}, N. and {Wilms}, J. and {Rothschild}, R.~E.},
        title = "{Two \raisebox{-0.5ex}\textasciitilde35 day clocks in Hercules X-1: evidence for neutron star free precession}",
      journal = {\aap},
         year = 2009,
        month = feb,
       volume = {494},
       number = {3},
        pages = {1025-1030},
          doi = {10.1051/0004-6361:200810743},
archivePrefix = {arXiv},
       eprint = {0811.4045},
 primaryClass = {astro-ph},
       adsurl = {https://ui.adsabs.harvard.edu/abs/2009A&A...494.1025S}
}

@ARTICLE{OGE85,
       author = {{Oegelman}, H. and {Kahabka}, P. and {Pietsch}, W. and {Truemper}, J. and {Voges}, W.},
        title = "{35-DAY Cycle and the Rotation Period of HERCULES-X-1}",
      journal = {\ssr},
         year = 1985,
        month = feb,
       volume = {40},
       number = {1-2},
        pages = {347-353},
          doi = {10.1007/BF00212907},
       adsurl = {https://ui.adsabs.harvard.edu/abs/1985SSRv...40..347O}
}

@ARTICLE{pringle_self-induced_1996,
       author = {{Pringle}, J.~E.},
        title = "{Self-induced warping of accretion discs}",
      journal = {\mnras},
         year = 1996,
        month = jul,
       volume = {281},
       number = {1},
        pages = {357-361},
          doi = {10.1093/mnras/281.1.357},
       adsurl = {https://ui.adsabs.harvard.edu/abs/1996MNRAS.281..357P}
}

@ARTICLE{ogilvie_precessing_2001,
       author = {{Ogilvie}, G.~I. and {Dubus}, G.},
        title = "{Precessing warped accretion discs in X-ray binaries}",
      journal = {\mnras},
         year = 2001,
        month = feb,
       volume = {320},
       number = {4},
        pages = {485-503},
          doi = {10.1046/j.1365-8711.2001.04011.x},
archivePrefix = {arXiv},
       eprint = {astro-ph/0009264},
 primaryClass = {astro-ph},
       adsurl = {https://ui.adsabs.harvard.edu/abs/2001MNRAS.320..485O}
}

@ARTICLE{whitehurst_superhumps_1991,
       author = {{Whitehurst}, Robert and {King}, Andrew},
        title = "{Superhumps, resonances and accretion discs.}",
      journal = {\mnras},
         year = 1991,
        month = mar,
       volume = {249},
        pages = {25},
          doi = {10.1093/mnras/249.1.25},
       adsurl = {https://ui.adsabs.harvard.edu/abs/1991MNRAS.249...25W}
}

@ARTICLE{maloney_radiation-driven_1996,
       author = {{Maloney}, Philip R. and {Begelman}, Mitchell C. and {Pringle}, J.~E.},
        title = "{Radiation-driven Warping: The Origin of WARPS and Precession in Accretion Disks}",
      journal = {\apj},
         year = 1996,
        month = nov,
       volume = {472},
        pages = {582},
          doi = {10.1086/178089},
archivePrefix = {arXiv},
       eprint = {astro-ph/9612110},
 primaryClass = {astro-ph},
       adsurl = {https://ui.adsabs.harvard.edu/abs/1996ApJ...472..582M}
}

@ARTICLE{petterson_35_1977,
       author = {{Petterson}, J.~A.},
        title = "{The 35 day cycle of the X-ray binary Hercules X-1.}",
      journal = {\apj},
         year = 1977,
        month = dec,
       volume = {218},
        pages = {783-791},
          doi = {10.1086/155735},
       adsurl = {https://ui.adsabs.harvard.edu/abs/1977ApJ...218..783P}
}

@ARTICLE{clarkson_II,
       author = {{Clarkson}, W.~I. and {Charles}, P.~A. and {Coe}, M.~J. and {Laycock}, S.},
        title = "{Long-term properties of accretion discs in X-ray binaries - II. Stability of radiation-driven warping}",
      journal = {\mnras},
         year = 2003,
        month = aug,
       volume = {343},
       number = {4},
        pages = {1213-1223},
          doi = {10.1046/j.1365-8711.2003.06761.x},
archivePrefix = {arXiv},
       eprint = {astro-ph/0304073},
 primaryClass = {astro-ph},
       adsurl = {https://ui.adsabs.harvard.edu/abs/2003MNRAS.343.1213C}
}

@ARTICLE{kotze_characterizing_2012,
       author = {{Kotze}, M.~M. and {Charles}, P.~A.},
        title = "{Characterizing X-ray binary long-term variability}",
      journal = {\mnras},
         year = 2012,
        month = feb,
       volume = {420},
       number = {2},
        pages = {1575-1589},
          doi = {10.1111/j.1365-2966.2011.20146.x},
archivePrefix = {arXiv},
       eprint = {1111.1296},
 primaryClass = {astro-ph.HE},
       adsurl = {https://ui.adsabs.harvard.edu/abs/2012MNRAS.420.1575K}
}

@ARTICLE{shakura2021nature,
       author = {{Shakura}, N.~I. and {Kolesnikov}, D.~A. and {Postnov}, K.~A.},
        title = "{On the Nature of the 35-Day Cycle in HZ Her/Her X-1}",
      journal = {Astronomy Reports},
         year = 2021,
        month = oct,
       volume = {65},
       number = {10},
        pages = {1039-1041},
          doi = {10.1134/S1063772921100334},
archivePrefix = {arXiv},
       eprint = {2103.01820},
 primaryClass = {astro-ph.HE},
       adsurl = {https://ui.adsabs.harvard.edu/abs/2021ARep...65.1039S}
}

@ARTICLE{staubert_variable_2013,
       author = {{Staubert}, R. and {Klochkov}, D. and {Vasco}, D. and {Postnov}, K. and {Shakura}, N. and {Wilms}, J. and {Rothschild}, R.~E.},
        title = "{Variable pulse profiles of Hercules X-1 repeating with the same irregular 35 d clock as the turn-ons}",
      journal = {\aap},
         year = 2013,
        month = feb,
       volume = {550},
          eid = {A110},
        pages = {A110},
          doi = {10.1051/0004-6361/201220316},
archivePrefix = {arXiv},
       eprint = {1212.5472},
 primaryClass = {astro-ph.HE},
       adsurl = {https://ui.adsabs.harvard.edu/abs/2013A&A...550A.110S}
}

@ARTICLE{kolesnikov2022evidence,
       author = {{Kolesnikov}, Dmitry and {Shakura}, Nikolai and {Postnov}, Konstantin},
        title = "{Evidence for neutron star triaxial free precession in Her X-1 from Fermi/GBM pulse period measurements}",
      journal = {\mnras},
         year = 2022,
        month = jul,
       volume = {513},
       number = {3},
        pages = {3359-3367},
          doi = {10.1093/mnras/stac1107},
archivePrefix = {arXiv},
       eprint = {2204.06408},
 primaryClass = {astro-ph.HE},
       adsurl = {https://ui.adsabs.harvard.edu/abs/2022MNRAS.513.3359K}
}

@ARTICLE{radhakrishnan1969magnetic,
       author = {{Radhakrishnan}, V. and {Cooke}, D.~J.},
        title = "{Magnetic Poles and the Polarization Structure of Pulsar Radiation}",
      journal = {\aplett},
         year = 1969,
        month = jan,
       volume = {3},
        pages = {225},
       adsurl = {https://ui.adsabs.harvard.edu/abs/1969ApL.....3..225R}
}

@ARTICLE{klochkov2009continuous,
       author = {{Klochkov}, D. and {Staubert}, R. and {Postnov}, K. and {Shakura}, N. and {Santangelo}, A.},
        title = "{Continuous monitoring of pulse period variations in Hercules X-1 using Swift/BAT}",
      journal = {\aap},
         year = 2009,
        month = nov,
       volume = {506},
       number = {3},
        pages = {1261-1267},
          doi = {10.1051/0004-6361/200912791},
archivePrefix = {arXiv},
       eprint = {0908.0053},
 primaryClass = {astro-ph.HE},
       adsurl = {https://ui.adsabs.harvard.edu/abs/2009A&A...506.1261K}
}

@ARTICLE{malacaria2020ups,
       author = {{Malacaria}, C. and {Jenke}, P. and {Roberts}, O.~J. and {Wilson-Hodge}, C.~A. and {Cleveland}, W.~H. and {Mailyan}, B. and {GBM Accreting Pulsars Program Team}},
        title = "{The Ups and Downs of Accreting X-Ray Pulsars: Decade-long Observations with the Fermi Gamma-Ray Burst Monitor}",
      journal = {\apj},
         year = 2020,
        month = jun,
       volume = {896},
       number = {1},
          eid = {90},
        pages = {90},
          doi = {10.3847/1538-4357/ab855c},
archivePrefix = {arXiv},
       eprint = {2004.00051},
 primaryClass = {astro-ph.HE},
       adsurl = {https://ui.adsabs.harvard.edu/abs/2020ApJ...896...90M}
}

@ARTICLE{BOY80,
       author = {{Boynton}, P.~E. and {Crosa}, L.~M. and {Deeter}, J.~E.},
        title = "{HER X-1/HZ HER : the 35 day clock.}",
      journal = {\apj},
         year = 1980,
        month = apr,
       volume = {237},
        pages = {169-174},
          doi = {10.1086/157856},
       adsurl = {https://ui.adsabs.harvard.edu/abs/1980ApJ...237..169B}
}

@ARTICLE{CRO80,
       author = {{Crosa}, L. and {Boynton}, P.~E.},
        title = "{Periodic mass transfer in HER X-1/ HZ Her.}",
      journal = {\apj},
         year = 1980,
        month = feb,
       volume = {235},
        pages = {999-1015},
          doi = {10.1086/157705},
       adsurl = {https://ui.adsabs.harvard.edu/abs/1980ApJ...235..999C}
}

@ARTICLE{THO83,
       author = {{Thomas}, H.~C. and {Africano}, J. and {Delgado}, A.~J. and {Schmidt}, H.~U.},
        title = "{Photometric observations. Is HZ Herculis getting darker ?}",
      journal = {\aap},
         year = 1983,
        month = oct,
       volume = {126},
        pages = {45-50},
       adsurl = {https://ui.adsabs.harvard.edu/abs/1983A&A...126...45T}
}

@INPROCEEDINGS{RIC82,
       author = {{Ricketts}, M.~J. and {Stanger}, V. and {Page}, C.~G.},
        title = "{HER X-1: Ariel-5 and Ariel-6 observations}",
    booktitle = {Accreting Neutron Stars},
         year = 1982,
       editor = {{Brinkmann}, W. and {Truemper}, J.},
        month = oct,
        pages = {100-105},
       adsurl = {https://ui.adsabs.harvard.edu/abs/1982ans..conf..100R}
}

@ARTICLE{JOSS76,
       author = {{Jones}, C. and {Forman}, W.},
        title = "{Uhuru observations of Hercules X-1 during the low state of the 35-day cycle.}",
      journal = {\apjl},
         year = 1976,
        month = nov,
       volume = {209},
        pages = {L131-L135},
          doi = {10.1086/182283},
       adsurl = {https://ui.adsabs.harvard.edu/abs/1976ApJ...209L.131J}
}

@misc{DAV77,
  author       = {Davison, P. J. N.},
  year         = {1977},
  howpublished = {IAU Circular 3039},
}

@ARTICLE{GOR81,
       author = {{Gorecki}, A. and {Levine}, A. and {Bautz}, M. and {Lang}, F. and {Primini}, F.~A. and {Lewin}, W.~H.~G. and {Baity}, W.~A. and {Gruber}, D.~E. and {Rothschild}, R.~E.},
        title = "{HEAO 1 observations of the long-term variability of HER X-1.}",
      journal = {\apj},
         year = 1982,
        month = may,
       volume = {256},
        pages = {234-237},
          doi = {10.1086/159900},
       adsurl = {https://ui.adsabs.harvard.edu/abs/1982ApJ...256..234G}
}

@ARTICLE{SOO90,
       author = {{Soong}, Y. and {Gruber}, D.~E. and {Peterson}, L.~E. and {Rothschild}, R.~E.},
        title = "{Temporal Behavior of Hercules X-1: The Long-Term Variability of Its Pulse Profile and the 35 Day X-Ray Intensity Modulation}",
      journal = {\apj},
         year = 1990,
        month = jan,
       volume = {348},
        pages = {634},
          doi = {10.1086/168271},
       adsurl = {https://ui.adsabs.harvard.edu/abs/1990ApJ...348..634S}
}

@ARTICLE{NAG84,
       author = {{Nagase}, F. and {Hayakawa}, S. and {Kii}, T. and {Sato}, N. and {Ikegami}, T. and {Kawai}, N. and {Makishima}, K. and {Matsuoka}, M. and {Mitani}, K. and {Murakami}, T. and {Oda}, M. and {Ohashi}, T. and {Tanaka}, Y. and {Kitamoto}, S.},
        title = "{Pulse-period changes of X-ray pulsars measured with Hakucho and Tenma.}",
      journal = {\pasj},
         year = 1984,
        month = jan,
       volume = {36},
        pages = {667-678},
       adsurl = {https://ui.adsabs.harvard.edu/abs/1984PASJ...36..667N}
}

@ARTICLE{DEL83,
       author = {{Delgado}, A.~J. and {Schmidt}, H.~U. and {Thomas}, H. -C.},
        title = "{HZ Herculis, still active.}",
      journal = {\aap},
         year = 1983,
        month = nov,
       volume = {127},
        pages = {L15-L16},
       adsurl = {https://ui.adsabs.harvard.edu/abs/1983A&A...127L..15D}
}

@article{DET93,
  author  = {Deeter, J. E. and Boynton, P. E. and Scott, M. D. and Miyamoto, S. and Kitamoto, S. and Nagase, F.},
  year    = {1993},
  title   = {},
  journal = {ApJ},
  note    = {unpublished}
}

\FloatBarrier
\appendix
\section{}

\begin{table*}[ht] 
\caption{Main-on peak times as reported in the literature and derived in this work (Sect.\ref{sec:Pdot_suporb_from_O_C}). Turn-on times compiled by B93 and STA06 have been shifted by $\sim$7 d to convert into main-on peak times. \label{tab:mainon_peak_times}}
{\scriptsize Note: References for the turn-on times are BOY80: \citet{BOY80}, CRO80: \citet{CRO80}, GIA73: \citet{herx1_suporb_discovery}, THO83: \citet{THO83}, STA83: \citet{STA83}, RIC82: \citet{RIC82}, JOSS76: \citet{JOSS76}, DAV77: \citet{DAV77}, GOR81: \citet{GOR81}, SOO90: \citet{SOO90}, NAG84: \citet{NAG84}, DEL83: \citet{DEL83}, DET93: \citet{DET93}, OGE85: \citet{OGE85}, and STA06: \citet{STA06}. Superscripts a, b, c, d, and e correspond to instruments \batse, \rxte, \swift, \fermi, and \maxi, respectively, used to determine the main-on peak time.}

\label{tab:observation_log}
\setlength{\tabcolsep}{12pt}
\resizebox{1.94\columnwidth}{!}{
\fontsize{6}{7}\selectfont
\begin{minipage}[t]{0.34\textwidth}
\begin{tabular}[t]{|lcc|}
\hline
Source & Cycle & Main-on peak (MJD)\\
\hline
B93 (BOY80) & 0 & 40805.170 $\pm$ 0.800 \\
B93 (CRO80) & 5 & 40980.070 $\pm$ 0.500 \\
B93 (BOY80) & 8 & 41084.470 $\pm$ 0.300 \\
B93 (GIA73) & 13 & 41261.970 $\pm$ 0.400 \\
B93 (CRO80) & 14 & 41297.394 $\pm$ 0.022 \\
B93 (CRO80) & 15 & 41333.050 $\pm$ 0.060 \\
B93 (CRO80) & 16 & 41368.740 $\pm$ 0.040 \\
B93 (CRO80) & 17 & 41404.370 $\pm$ 0.040 \\
B93 (CRO80) & 18 & 41439.462 $\pm$ 0.070 \\
B93 (CRO80) & 19 & 41474.230 $\pm$ 0.050 \\
B93 (CRO80) & 20 & 41509.019 $\pm$ 0.038 \\
B93 (CRO80) & 21 & 41543.860 $\pm$ 0.040 \\
B93 (CRO80) & 22 & 41579.478 $\pm$ 0.023 \\
B93 (CRO80) & 23 & 41613.460 $\pm$ 0.040 \\
B93 (CRO80) & 24 & 41647.594 $\pm$ 0.033 \\
B93 (CRO80) & 25 & 41682.374 $\pm$ 0.030 \\
B93 (CRO80) & 26 & 41716.560 $\pm$ 0.070 \\
B93 (CRO80) & 27 & 41750.590 $\pm$ 0.040 \\
B93 (THO83) & 28 & 41784.490 $\pm$ 0.500 \\
B93 (THO83) & 29 & 41820.180 $\pm$ 0.500 \\
B93 (THO83) & 30 & 41855.880 $\pm$ 0.500 \\
B93 (THO83) & 31 & 41891.600 $\pm$ 0.500 \\
B93 (STA83) & 32 & 41925.961 $\pm$ 0.410 \\
B93 (THO83) & 33 & 41961.310 $\pm$ 0.500 \\
B93 (CRO80) & 39 & 42171.114 $\pm$ 0.017 \\
B93 (BOY80) & 41 & 42240.170 $\pm$ 0.300 \\
B93 (STA83) & 45 & 42382.972 $\pm$ 0.100 \\
B93 (STA83) & 46 & 42417.813 $\pm$ 0.100 \\
B93 (STA83) & 47 & 42453.779 $\pm$ 0.100 \\
B93 (CRO80) & 48 & 42487.510 $\pm$ 0.070 \\
B93 (CRO80) & 49 & 42521.470 $\pm$ 0.020 \\
B93 (BOY80) & 51 & 42590.170 $\pm$ 0.200 \\
B93 (CRO80) & 53 & 42660.055 $\pm$ 0.035 \\
B93 (RIC82) & 58 & 42834.470 $\pm$ 0.300 \\
B93 (RIC82) & 61 & 42939.670 $\pm$ 0.200 \\
B93 (JOSS76) & 62 & 42975.380 $\pm$ 0.045 \\
B93 (DAV77) & 68 & 43185.500 $\pm$ 0.042 \\
B93 (THO83) & 73 & 43357.960 $\pm$ 0.500 \\
B93 (STA83) & 74 & 43393.570 $\pm$ 0.100 \\
B93 (GOR81) & 79 & 43567.970 $\pm$ 0.100 \\
B93 (GOR81) & 84 & 43740.270 $\pm$ 0.300 \\
B93 (SOO90) & 85 & 43776.370 $\pm$ 0.500 \\
B93 (RIC82) & 87 & 43845.770 $\pm$ 0.200 \\
B93 (RIC82) & 88 & 43880.870 $\pm$ 0.200 \\
B93 (THO83) & 92 & 44022.720 $\pm$ 0.500 \\
B93 (RIC82) & 99 & 44268.370 $\pm$ 0.200 \\
B93 (THO83) & 103 & 44407.839 $\pm$ 0.500 \\
B93 (THO83) & 112 & 44716.382 $\pm$ 0.500 \\
B93 (THO83) & 113 & 44751.280 $\pm$ 0.500 \\
B93 (RIC82) & 114 & 44786.870 $\pm$ 1.000 \\
B93 (NAG84) & 124 & 45134.870 $\pm$ 1.000 \\
B93 (THO83) & 125 & 45171.210 $\pm$ 0.500 \\
B93 (DEL83) & 134 & 45479.370 $\pm$ 0.500 \\
B93 (OGE85) & 142 & 45760.370 $\pm$ 1.000 \\
B93 (OGE85) & 143 & 45795.370 $\pm$ 0.500 \\
B93 (OGE85) & 155 & 46215.670 $\pm$ 1.000 \\
B93 (DET93) & 189 & 47406.170 $\pm$ 1.000 \\
B93 (DET93) & 196 & 47650.470 $\pm$ 0.100 \\
B93 (DET93) & 220 & 48485.670 $\pm$ 0.200 \\
STA06 & 221 & 48515.148 $\pm$ 0.500 \\
STA06 & 222 & 48550.941 $\pm$ 0.500 \\
STA06 & 223 & 48585.735 $\pm$ 0.500 \\
STA06 & 224 & 48619.749 $\pm$ 0.500 \\
STA06 & 225 & 48653.742 $\pm$ 0.500 \\
STA06 & 226 & 48687.636 $\pm$ 0.500 \\
STA06 & 227 & 48723.229 $\pm$ 0.500 \\
STA06 & 228 & 48758.243 $\pm$ 0.500 \\
STA06 & 229 & 48792.236 $\pm$ 0.500 \\
STA06 & 230 & 48826.050 $\pm$ 0.500 \\
STA06 & 231 & 48860.044 $\pm$ 0.500 \\
STA06 & 232 & 48894.338 $\pm$ 0.500 \\
STA06 & 233 & 48929.131 $\pm$ 0.500 \\
STA06 & 234 & 48963.845 $\pm$ 0.500 \\
STA06 & 236 & 49033.832 $\pm$ 0.500 \\
STA06 & 237 & 49068.765 $\pm$ 0.500 \\
STA06 & 238 & 49104.238 $\pm$ 0.500 \\
STA06 & 239 & 49139.691 $\pm$ 0.500 \\
STA06 & 240 & 49174.845 $\pm$ 0.500 \\
STA06 & 241 & 49208.739 $\pm$ 0.500 \\
STA06 & 242 & 49242.533 $\pm$ 0.500 \\
STA06 & 244 & 49311.480 $\pm$ 0.500 \\
STA06 & 245 & 49346.233 $\pm$ 0.500 \\
STA06 & 246 & 49381.167 $\pm$ 0.500 \\
STA06 & 247 & 49415.920 $\pm$ 0.500 \\
STA06 & 248 & 49450.993 $\pm$ 0.500 \\
STA06 & 249 & 49485.647 $\pm$ 0.500 \\
STA06 & 250 & 49521.340 $\pm$ 0.500 \\
STA06 & 251 & 49555.634 $\pm$ 0.500 \\
STA06 & 252 & 49589.747 $\pm$ 0.500 \\
STA06 & 253 & 49624.641 $\pm$ 0.500 \\
STA06 & 254 & 49659.934 $\pm$ 0.500 \\
STA06 & 255 & 49694.827 $\pm$ 0.500 \\
STA06 & 256 & 49729.641 $\pm$ 0.500 \\
STA06 & 257 & 49764.734 $\pm$ 0.500 \\
STA06 & 258 & 49798.528 $\pm$ 0.500 \\
STA06 & 259 & 49834.441 $\pm$ 0.500 \\
STA06 & 260 & 49869.035 $\pm$ 0.500 \\
\hline
\end{tabular}
\end{minipage}

\begin{minipage}[t]{0.34\textwidth}
\begin{tabular}[t]{|lcc|}
\hline
Source & Cycle & Main-on peak (MJD)\\
\hline
This Work$^{\mathrm{a}}$ & 261 & 49904.438 $\pm$ 0.214 \\
This Work$^{\mathrm{a}}$ & 262 & 49938.334 $\pm$ 0.365 \\
This Work$^{\mathrm{a}}$ & 263 & 49974.482 $\pm$ 0.473 \\
This Work$^{\mathrm{a}}$ & 264 & 50009.956 $\pm$ 0.347 \\
This Work$^{\mathrm{a}}$ & 265 & 50046.165 $\pm$ 0.265 \\
This Work$^{\mathrm{a,b}}$ & 266 & 50080.591 $\pm$ 0.204 \\
This Work$^{\mathrm{a}}$ & 267 & 50114.506 $\pm$ 0.124 \\
This Work$^{\mathrm{a,b}}$ & 268 & 50149.772 $\pm$ 0.172 \\
This Work$^{\mathrm{a,b}}$ & 269 & 50184.100 $\pm$ 0.157 \\
This Work$^{\mathrm{a,b}}$ & 270 & 50219.885 $\pm$ 0.141 \\
This Work$^{\mathrm{a,b}}$ & 271 & 50255.101 $\pm$ 0.220 \\
This Work$^{\mathrm{a,b}}$ & 272 & 50289.119 $\pm$ 0.141 \\
This Work$^{\mathrm{a,b}}$ & 273 & 50324.071 $\pm$ 0.143 \\
This Work$^{\mathrm{a,b}}$ & 274 & 50359.070 $\pm$ 0.164 \\
This Work$^{\mathrm{a,b}}$ & 275 & 50393.721 $\pm$ 0.119 \\
This Work$^{\mathrm{a,b}}$ & 276 & 50428.779 $\pm$ 0.130 \\
This Work$^{\mathrm{a,b}}$ & 277 & 50464.058 $\pm$ 0.147 \\
This Work$^{\mathrm{a,b}}$ & 278 & 50499.933 $\pm$ 0.159 \\
This Work$^{\mathrm{a,b}}$ & 279 & 50534.611 $\pm$ 0.190 \\
This Work$^{\mathrm{a,b}}$ & 280 & 50569.421 $\pm$ 0.148 \\
This Work$^{\mathrm{a,b}}$ & 281 & 50604.717 $\pm$ 0.200 \\
This Work$^{\mathrm{a,b}}$ & 282 & 50639.329 $\pm$ 0.120 \\
This Work$^{\mathrm{a,b}}$ & 283 & 50674.108 $\pm$ 0.227 \\
This Work$^{\mathrm{a,b}}$ & 284 & 50708.397 $\pm$ 0.140 \\
This Work$^{\mathrm{a,b}}$ & 285 & 50741.789 $\pm$ 0.172 \\
This Work$^{\mathrm{a,b}}$ & 286 & 50775.952 $\pm$ 0.237 \\
This Work$^{\mathrm{a,b}}$ & 287 & 50810.907 $\pm$ 0.123 \\
This Work$^{\mathrm{a,b}}$ & 288 & 50845.782 $\pm$ 0.114 \\
This Work$^{\mathrm{a,b}}$ & 289 & 50880.301 $\pm$ 0.158 \\
This Work$^{\mathrm{a,b}}$ & 290 & 50916.004 $\pm$ 0.101 \\
This Work$^{\mathrm{a,b}}$ & 291 & 50950.842 $\pm$ 0.325 \\
This Work$^{\mathrm{a,b}}$ & 292 & 50985.798 $\pm$ 0.091 \\
This Work$^{\mathrm{a,b}}$ & 293 & 51019.426 $\pm$ 0.205 \\
This Work$^{\mathrm{a,b}}$ & 294 & 51053.570 $\pm$ 0.193 \\
This Work$^{\mathrm{a,b}}$ & 295 & 51089.099 $\pm$ 0.097 \\
This Work$^{\mathrm{a,b}}$ & 296 & 51123.425 $\pm$ 0.141 \\
This Work$^{\mathrm{a,b}}$ & 297 & 51156.972 $\pm$ 0.140 \\
This Work$^{\mathrm{a,b}}$ & 298 & 51191.800 $\pm$ 0.153 \\
This Work$^{\mathrm{b}}$ & 316 & 51829.561 $\pm$ 0.271 \\
This Work$^{\mathrm{b}}$ & 317 & 51863.609 $\pm$ 0.588 \\
This Work$^{\mathrm{b}}$ & 318 & 51898.980 $\pm$ 0.552 \\
This Work$^{\mathrm{b}}$ & 319 & 51932.781 $\pm$ 0.183 \\
This Work$^{\mathrm{b}}$ & 320 & 51966.805 $\pm$ 0.159 \\
This Work$^{\mathrm{b}}$ & 321 & 52002.755 $\pm$ 0.902 \\
This Work$^{\mathrm{b}}$ & 322 & 52035.909 $\pm$ 0.231 \\
This Work$^{\mathrm{b}}$ & 323 & 52071.674 $\pm$ 0.189 \\
This Work$^{\mathrm{b}}$ & 324 & 52106.532 $\pm$ 0.328 \\
This Work$^{\mathrm{b}}$ & 325 & 52141.455 $\pm$ 0.149 \\
This Work$^{\mathrm{b}}$ & 326 & 52177.786 $\pm$ 0.263 \\
This Work$^{\mathrm{b}}$ & 327 & 52212.713 $\pm$ 0.248 \\
This Work$^{\mathrm{b}}$ & 329 & 52282.512 $\pm$ 0.317 \\
This Work$^{\mathrm{b}}$ & 330 & 52317.742 $\pm$ 0.202 \\
This Work$^{\mathrm{b}}$ & 331 & 52352.522 $\pm$ 0.212 \\
This Work$^{\mathrm{b}}$ & 332 & 52387.827 $\pm$ 0.242 \\
This Work$^{\mathrm{b}}$ & 333 & 52422.487 $\pm$ 0.184 \\
This Work$^{\mathrm{b}}$ & 334 & 52458.145 $\pm$ 0.210 \\
This Work$^{\mathrm{b}}$ & 335 & 52493.217 $\pm$ 0.188 \\
This Work$^{\mathrm{b}}$ & 336 & 52528.408 $\pm$ 0.153 \\
This Work$^{\mathrm{b}}$ & 337 & 52563.295 $\pm$ 0.289 \\
This Work$^{\mathrm{b}}$ & 338 & 52599.668 $\pm$ 0.583 \\
This Work$^{\mathrm{b}}$ & 339 & 52632.624 $\pm$ 0.231 \\
This Work$^{\mathrm{b}}$ & 340 & 52670.375 $\pm$ 0.431 \\
This Work$^{\mathrm{b}}$ & 341 & 52704.043 $\pm$ 0.223 \\
This Work$^{\mathrm{b}}$ & 342 & 52739.112 $\pm$ 0.188 \\
This Work$^{\mathrm{b}}$ & 343 & 52774.686 $\pm$ 0.282 \\
This Work$^{\mathrm{b}}$ & 344 & 52808.811 $\pm$ 0.366 \\
This Work$^{\mathrm{b}}$ & 345 & 52845.778 $\pm$ 0.237 \\
This Work$^{\mathrm{b}}$ & 346 & 52879.635 $\pm$ 0.174 \\
This Work$^{\mathrm{b}}$ & 347 & 52912.748 $\pm$ 0.356 \\
This Work$^{\mathrm{b}}$ & 348 & 52947.667 $\pm$ 0.275 \\
This Work$^{\mathrm{b}}$ & 350 & 53017.806 $\pm$ 0.642 \\
This Work$^{\mathrm{b}}$ & 354 & 53162.265 $\pm$ 0.404 \\
This Work$^{\mathrm{b}}$ & 355 & 53196.632 $\pm$ 0.309 \\
This Work$^{\mathrm{b}}$ & 356 & 53231.377 $\pm$ 0.270 \\
This Work$^{\mathrm{b}}$ & 357 & 53266.342 $\pm$ 0.290 \\
This Work$^{\mathrm{b}}$ & 358 & 53300.175 $\pm$ 0.138 \\
This Work$^{\mathrm{b}}$ & 359 & 53334.130 $\pm$ 0.712 \\
This Work$^{\mathrm{b}}$ & 360 & 53368.980 $\pm$ 0.358 \\
This Work$^{\mathrm{b,c}}$ & 361 & 53403.380 $\pm$ 0.244 \\
This Work$^{\mathrm{b,c}}$ & 362 & 53438.503 $\pm$ 0.217 \\
This Work$^{\mathrm{b,c}}$ & 363 & 53472.141 $\pm$ 0.197 \\
This Work$^{\mathrm{b,c}}$ & 364 & 53508.041 $\pm$ 0.202 \\
This Work$^{\mathrm{b,c}}$ & 365 & 53543.856 $\pm$ 0.197 \\
\hline 
\end{tabular}
\end{minipage}

\begin{minipage}[t]{0.34\textwidth}
\begin{tabular}[t]{|lcc|}
\hline
Source & Cycle & Main-on peak (MJD)\\
\hline
This Work$^{\mathrm{b,c}}$ & 366 & 53579.219 $\pm$ 0.348 \\
This Work$^{\mathrm{b,c}}$ & 367 & 53614.339 $\pm$ 0.286 \\
This Work$^{\mathrm{b,c}}$ & 368 & 53648.330 $\pm$ 0.087 \\
This Work$^{\mathrm{b,c}}$ & 369 & 53683.631 $\pm$ 0.171 \\
This Work$^{\mathrm{b,c}}$ & 370 & 53717.535 $\pm$ 0.139 \\
This Work$^{\mathrm{b,c}}$ & 371 & 53753.122 $\pm$ 0.158 \\
This Work$^{\mathrm{b,c}}$ & 372 & 53787.670 $\pm$ 0.454 \\
This Work$^{\mathrm{b,c}}$ & 373 & 53821.649 $\pm$ 0.176 \\
This Work$^{\mathrm{b,c}}$ & 374 & 53857.064 $\pm$ 0.088 \\
This Work$^{\mathrm{b,c}}$ & 375 & 53891.423 $\pm$ 0.580 \\
This Work$^{\mathrm{b,c}}$ & 376 & 53926.429 $\pm$ 0.180 \\
This Work$^{\mathrm{b,c}}$ & 377 & 53962.404 $\pm$ 0.167 \\
This Work$^{\mathrm{b,c}}$ & 378 & 53997.975 $\pm$ 0.184 \\
This Work$^{\mathrm{b,c}}$ & 379 & 54033.953 $\pm$ 0.391 \\
This Work$^{\mathrm{b,c}}$ & 380 & 54067.374 $\pm$ 0.144 \\
This Work$^{\mathrm{b,c}}$ & 381 & 54102.560 $\pm$ 0.375 \\
This Work$^{\mathrm{b,c}}$ & 382 & 54136.579 $\pm$ 0.199 \\
This Work$^{\mathrm{b,c}}$ & 383 & 54171.355 $\pm$ 0.240 \\
This Work$^{\mathrm{b,c}}$ & 384 & 54205.920 $\pm$ 0.161 \\
This Work$^{\mathrm{b,c}}$ & 385 & 54240.294 $\pm$ 0.248 \\
This Work$^{\mathrm{b,c}}$ & 386 & 54274.429 $\pm$ 0.225 \\
This Work$^{\mathrm{b,c}}$ & 387 & 54309.914 $\pm$ 0.172 \\
This Work$^{\mathrm{b,c}}$ & 388 & 54344.340 $\pm$ 0.288 \\
This Work$^{\mathrm{b,c}}$ & 389 & 54378.111 $\pm$ 0.244 \\
This Work$^{\mathrm{b,c}}$ & 390 & 54411.867 $\pm$ 0.133 \\
This Work$^{\mathrm{b,c}}$ & 391 & 54447.159 $\pm$ 0.131 \\
This Work$^{\mathrm{b,c}}$ & 392 & 54482.238 $\pm$ 0.175 \\
This Work$^{\mathrm{b,c}}$ & 393 & 54516.154 $\pm$ 0.222 \\
This Work$^{\mathrm{b,c}}$ & 394 & 54551.172 $\pm$ 0.192 \\
This Work$^{\mathrm{b,c}}$ & 395 & 54586.395 $\pm$ 0.231 \\
This Work$^{\mathrm{b,c}}$ & 396 & 54620.474 $\pm$ 0.185 \\
This Work$^{\mathrm{b,c}}$ & 397 & 54657.168 $\pm$ 0.206 \\
This Work$^{\mathrm{b,c}}$ & 398 & 54691.599 $\pm$ 0.178 \\
This Work$^{\mathrm{b,c,d}}$ & 399 & 54725.545 $\pm$ 0.045 \\
This Work$^{\mathrm{b,c,d}}$ & 400 & 54761.271 $\pm$ 0.239 \\
This Work$^{\mathrm{b,c,d}}$ & 401 & 54795.249 $\pm$ 0.147 \\
This Work$^{\mathrm{b,c,d}}$ & 402 & 54830.789 $\pm$ 0.093 \\
This Work$^{\mathrm{b,c,d}}$ & 403 & 54865.508 $\pm$ 0.064 \\
This Work$^{\mathrm{b,c,d}}$ & 404 & 54900.064 $\pm$ 0.240 \\
This Work$^{\mathrm{b,c,d}}$ & 405 & 54935.816 $\pm$ 0.103 \\
This Work$^{\mathrm{b,c,d}}$ & 406 & 54971.548 $\pm$ 0.086 \\
This Work$^{\mathrm{b,c,d}}$ & 407 & 55006.375 $\pm$ 0.144 \\
This Work$^{\mathrm{b,c,d}}$ & 408 & 55041.651 $\pm$ 0.061 \\
This Work$^{\mathrm{b,c,d}}$ & 409 & 55076.341 $\pm$ 0.060 \\
This Work$^{\mathrm{b,c,d,e}}$ & 410 & 55111.565 $\pm$ 0.102 \\
This Work$^{\mathrm{b,c,d,e}}$ & 411 & 55146.940 $\pm$ 0.081 \\
This Work$^{\mathrm{b,c,d,e}}$ & 412 & 55182.000 $\pm$ 0.087 \\
This Work$^{\mathrm{b,c,d,e}}$ & 413 & 55217.382 $\pm$ 0.060 \\
This Work$^{\mathrm{b,c,d,e}}$ & 414 & 55252.087 $\pm$ 0.052 \\
This Work$^{\mathrm{b,c,d,e}}$ & 415 & 55286.544 $\pm$ 0.059 \\
This Work$^{\mathrm{b,c,d,e}}$ & 416 & 55321.871 $\pm$ 0.051 \\
This Work$^{\mathrm{b,c,d,e}}$ & 417 & 55357.489 $\pm$ 0.078 \\
This Work$^{\mathrm{b,c,d,e}}$ & 418 & 55392.905 $\pm$ 0.060 \\
This Work$^{\mathrm{b,c,d,e}}$ & 419 & 55428.216 $\pm$ 0.086 \\
This Work$^{\mathrm{b,c,d,e}}$ & 420 & 55463.368 $\pm$ 0.068 \\
This Work$^{\mathrm{b,c,d,e}}$ & 421 & 55497.645 $\pm$ 0.114 \\
This Work$^{\mathrm{b,c,d,e}}$ & 422 & 55531.528 $\pm$ 0.038 \\
This Work$^{\mathrm{b,c,d,e}}$ & 423 & 55566.634 $\pm$ 0.058 \\
This Work$^{\mathrm{b,c,d,e}}$ & 424 & 55600.827 $\pm$ 0.069 \\
This Work$^{\mathrm{b,c,d,e}}$ & 425 & 55635.831 $\pm$ 0.080 \\
This Work$^{\mathrm{b,c,d,e}}$ & 426 & 55670.636 $\pm$ 0.126 \\
This Work$^{\mathrm{b,c,d,e}}$ & 427 & 55705.643 $\pm$ 0.087 \\
This Work$^{\mathrm{b,c,d,e}}$ & 428 & 55740.022 $\pm$ 0.103 \\
This Work$^{\mathrm{b,c,d,e}}$ & 429 & 55777.034 $\pm$ 0.164 \\
This Work$^{\mathrm{b,c,d,e}}$ & 430 & 55811.488 $\pm$ 0.116 \\
This Work$^{\mathrm{c,d,e}}$ & 431 & 55846.141 $\pm$ 0.098 \\
This Work$^{\mathrm{c,d,e}}$ & 432 & 55880.712 $\pm$ 0.039 \\
This Work$^{\mathrm{b,c,d,e}}$ & 433 & 55915.843 $\pm$ 0.128 \\
This Work$^{\mathrm{c,d,e}}$ & 434 & 55951.594 $\pm$ 0.056 \\
This Work$^{\mathrm{c,d,e}}$ & 435 & 55985.662 $\pm$ 0.081 \\
This Work$^{\mathrm{c,d,e}}$ & 436 & 56019.900 $\pm$ 0.113 \\
This Work$^{\mathrm{c,d,e}}$ & 437 & 56053.735 $\pm$ 0.119 \\
This Work$^{\mathrm{c,d,e}}$ & 438 & 56088.736 $\pm$ 0.102 \\
This Work$^{\mathrm{c,d,e}}$ & 439 & 56124.293 $\pm$ 0.117 \\
This Work$^{\mathrm{c,d,e}}$ & 440 & 56157.533 $\pm$ 0.098 \\
This Work$^{\mathrm{c,d,e}}$ & 441 & 56193.387 $\pm$ 0.074 \\
This Work$^{\mathrm{c,d,e}}$ & 442 & 56228.718 $\pm$ 0.086 \\
This Work$^{\mathrm{c,d,e}}$ & 443 & 56264.190 $\pm$ 0.114 \\
This Work$^{\mathrm{c,d,e}}$ & 444 & 56298.455 $\pm$ 0.054 \\
This Work$^{\mathrm{c,d,e}}$ & 445 & 56332.920 $\pm$ 0.063 \\
This Work$^{\mathrm{c,d,e}}$ & 446 & 56367.811 $\pm$ 0.096 \\
This Work$^{\mathrm{c,d,e}}$ & 447 & 56402.908 $\pm$ 0.103 \\
\hline 
\end{tabular}
\end{minipage}
}
\end{table*}

\begin{table*} 
\centering
\caption*{Table A.1: Continued.}
\label{tab:observation_log}
\setlength{\tabcolsep}{11pt}
\resizebox{1.94\columnwidth}{!}{
\fontsize{6}{7}\selectfont
\begin{minipage}[t]{0.34\textwidth}
\begin{tabular}[t]{|lcc|}
\hline
Source & Cycle & Main-on peak (MJD)\\
\hline

This Work$^{\mathrm{c,d,e}}$ & 448 & 56438.337 $\pm$ 0.080 \\
This Work$^{\mathrm{c,d,e}}$ & 449 & 56472.918 $\pm$ 0.073 \\
This Work$^{\mathrm{c,d,e}}$ & 450 & 56507.582 $\pm$ 0.063 \\
This Work$^{\mathrm{c,d,e}}$ & 451 & 56541.394 $\pm$ 0.064 \\
This Work$^{\mathrm{c,d,e}}$ & 452 & 56575.428 $\pm$ 0.048 \\
This Work$^{\mathrm{c,d,e}}$ & 453 & 56609.547 $\pm$ 0.061 \\
This Work$^{\mathrm{c,d,e}}$ & 454 & 56645.304 $\pm$ 0.127 \\
This Work$^{\mathrm{c,d,e}}$ & 455 & 56679.586 $\pm$ 0.058 \\
This Work$^{\mathrm{c,d,e}}$ & 456 & 56715.053 $\pm$ 0.152 \\
This Work$^{\mathrm{c,d,e}}$ & 457 & 56749.938 $\pm$ 0.053 \\
This Work$^{\mathrm{c,d,e}}$ & 458 & 56785.313 $\pm$ 0.108 \\
This Work$^{\mathrm{c,d,e}}$ & 459 & 56820.715 $\pm$ 0.094 \\
This Work$^{\mathrm{c,d,e}}$ & 460 & 56856.107 $\pm$ 0.098 \\
This Work$^{\mathrm{c,d,e}}$ & 461 & 56890.197 $\pm$ 0.137 \\
This Work$^{\mathrm{c,d,e}}$ & 462 & 56924.846 $\pm$ 0.059 \\
This Work$^{\mathrm{c,d,e}}$ & 463 & 56959.658 $\pm$ 0.077 \\
This Work$^{\mathrm{c,d,e}}$ & 464 & 56994.363 $\pm$ 0.059 \\
This Work$^{\mathrm{c,d,e}}$ & 465 & 57029.698 $\pm$ 0.097 \\
This Work$^{\mathrm{c,d,e}}$ & 466 & 57063.338 $\pm$ 0.154 \\
This Work$^{\mathrm{c,d,e}}$ & 467 & 57097.468 $\pm$ 0.093 \\
This Work$^{\mathrm{c,d,e}}$ & 468 & 57133.782 $\pm$ 0.073 \\
This Work$^{\mathrm{c,d,e}}$ & 469 & 57168.921 $\pm$ 0.048 \\
This Work$^{\mathrm{c,d,e}}$ & 470 & 57203.615 $\pm$ 0.087 \\
This Work$^{\mathrm{c,d,e}}$ & 471 & 57236.655 $\pm$ 0.065 \\
This Work$^{\mathrm{c,d,e}}$ & 472 & 57270.075 $\pm$ 0.067 \\
This Work$^{\mathrm{c,d,e}}$ & 473 & 57305.279 $\pm$ 0.053 \\
This Work$^{\mathrm{c,d,e}}$ & 474 & 57339.469 $\pm$ 0.290 \\
This Work$^{\mathrm{c,d,e}}$ & 475 & 57375.012 $\pm$ 0.069 \\
This Work$^{\mathrm{c,d,e}}$ & 476 & 57409.965 $\pm$ 0.051 \\
This Work$^{\mathrm{c,d,e}}$ & 477 & 57445.057 $\pm$ 0.073 \\
This Work$^{\mathrm{c,d,e}}$ & 478 & 57480.555 $\pm$ 0.065 \\
This Work$^{\mathrm{c,d,e}}$ & 479 & 57515.810 $\pm$ 0.093 \\
This Work$^{\mathrm{c,d,e}}$ & 480 & 57550.626 $\pm$ 0.106 \\
This Work$^{\mathrm{c,d,e}}$ & 481 & 57586.238 $\pm$ 0.128 \\
This Work$^{\mathrm{c,d,e}}$ & 482 & 57621.963 $\pm$ 0.120 \\
This Work$^{\mathrm{c,d,e}}$ & 483 & 57657.726 $\pm$ 0.073 \\
This Work$^{\mathrm{c,d,e}}$ & 484 & 57693.064 $\pm$ 0.083 \\
This Work$^{\mathrm{c,d,e}}$ & 485 & 57727.688 $\pm$ 0.078 \\
This Work$^{\mathrm{c,d,e}}$ & 486 & 57763.565 $\pm$ 0.116 \\
This Work$^{\mathrm{c,d,e}}$ & 487 & 57798.276 $\pm$ 0.060 \\
This Work$^{\mathrm{c,d,e}}$ & 488 & 57832.679 $\pm$ 0.057 \\
This Work$^{\mathrm{c,d,e}}$ & 489 & 57866.607 $\pm$ 0.091 \\
This Work$^{\mathrm{c,d,e}}$ & 490 & 57900.709 $\pm$ 0.052 \\
This Work$^{\mathrm{c,d,e}}$ & 491 & 57935.261 $\pm$ 0.078 \\
This Work$^{\mathrm{c,d,e}}$ & 492 & 57969.372 $\pm$ 0.075 \\
This Work$^{\mathrm{c,d,e}}$ & 493 & 58002.641 $\pm$ 0.050 \\
This Work$^{\mathrm{c,d,e}}$ & 494 & 58036.994 $\pm$ 0.049 \\
This Work$^{\mathrm{c,d,e}}$ & 495 & 58071.671 $\pm$ 0.098 \\
This Work$^{\mathrm{c,d,e}}$ & 496 & 58106.145 $\pm$ 0.076 \\
This Work$^{\mathrm{c,d,e}}$ & 497 & 58141.288 $\pm$ 0.136 \\
This Work$^{\mathrm{c,d,e}}$ & 498 & 58175.191 $\pm$ 0.083 \\
This Work$^{\mathrm{c,d,e}}$ & 499 & 58207.917 $\pm$ 0.123 \\
This Work$^{\mathrm{c,d,e}}$ & 500 & 58242.528 $\pm$ 0.051 \\
This Work$^{\mathrm{c,d,e}}$ & 501 & 58277.619 $\pm$ 0.055 \\
This Work$^{\mathrm{c,d,e}}$ & 502 & 58311.947 $\pm$ 0.043 \\
This Work$^{\mathrm{c,d,e}}$ & 503 & 58346.482 $\pm$ 0.074 \\
This Work$^{\mathrm{c,d,e}}$ & 504 & 58380.707 $\pm$ 0.066 \\
This Work$^{\mathrm{c,d,e}}$ & 505 & 58416.229 $\pm$ 0.110 \\
This Work$^{\mathrm{c,d,e}}$ & 506 & 58450.620 $\pm$ 0.085 \\
This Work$^{\mathrm{c,d,e}}$ & 507 & 58486.740 $\pm$ 0.087 \\
This Work$^{\mathrm{c,d,e}}$ & 508 & 58523.811 $\pm$ 0.083 \\
This Work$^{\mathrm{c,d,e}}$ & 509 & 58555.073 $\pm$ 0.099 \\
This Work$^{\mathrm{c,d,e}}$ & 510 & 58589.110 $\pm$ 0.077 \\
This Work$^{\mathrm{c,d,e}}$ & 511 & 58623.463 $\pm$ 0.061 \\
This Work$^{\mathrm{c,d,e}}$ & 512 & 58657.657 $\pm$ 0.116 \\
This Work$^{\mathrm{c,d,e}}$ & 513 & 58692.379 $\pm$ 0.052 \\
This Work$^{\mathrm{c,d,e}}$ & 514 & 58727.187 $\pm$ 0.137 \\
This Work$^{\mathrm{c,d,e}}$ & 515 & 58762.371 $\pm$ 0.061 \\
This Work$^{\mathrm{c,d,e}}$ & 516 & 58797.275 $\pm$ 0.139 \\
This Work$^{\mathrm{c,d,e}}$ & 517 & 58832.889 $\pm$ 0.110 \\
This Work$^{\mathrm{c,d,e}}$ & 518 & 58867.862 $\pm$ 0.087 \\
This Work$^{\mathrm{c,d,e}}$ & 519 & 58902.851 $\pm$ 0.085 \\
This Work$^{\mathrm{c,d,e}}$ & 520 & 58938.113 $\pm$ 0.065 \\
This Work$^{\mathrm{c,d,e}}$ & 521 & 58972.611 $\pm$ 0.088 \\
This Work$^{\mathrm{c,d,e}}$ & 522 & 59007.076 $\pm$ 0.095 \\
This Work$^{\mathrm{c,d,e}}$ & 523 & 59040.990 $\pm$ 0.061 \\
This Work$^{\mathrm{c,d,e}}$ & 524 & 59075.738 $\pm$ 0.081 \\
This Work$^{\mathrm{c,d,e}}$ & 525 & 59110.570 $\pm$ 0.071 \\
This Work$^{\mathrm{c,d,e}}$ & 526 & 59144.189 $\pm$ 0.047 \\
This Work$^{\mathrm{c,d,e}}$ & 527 & 59179.064 $\pm$ 0.176 \\
This Work$^{\mathrm{c,d,e}}$ & 528 & 59213.798 $\pm$ 0.067 \\
This Work$^{\mathrm{c,d,e}}$ & 529 & 59248.476 $\pm$ 0.079 \\
This Work$^{\mathrm{c,d,e}}$ & 530 & 59282.834 $\pm$ 0.067 \\
This Work$^{\mathrm{c,d,e}}$ & 531 & 59318.331 $\pm$ 0.062 \\
This Work$^{\mathrm{c,d,e}}$ & 532 & 59352.933 $\pm$ 0.079 \\
This Work$^{\mathrm{c,d,e}}$ & 533 & 59388.813 $\pm$ 0.175 \\
This Work$^{\mathrm{c,d,e}}$ & 534 & 59423.252 $\pm$ 0.065 \\
This Work$^{\mathrm{c,d,e}}$ & 535 & 59457.885 $\pm$ 0.047 \\
This Work$^{\mathrm{c,d,e}}$ & 536 & 59493.702 $\pm$ 0.109 \\
\hline
\end{tabular}
\end{minipage}

\begin{minipage}[t]{0.34\textwidth}
\begin{tabular}[t]{|lcc|}
\hline
Source & Cycle & Main-on peak (MJD)\\
\hline
This Work$^{\mathrm{c,d,e}}$ & 537 & 59527.624 $\pm$ 0.143 \\
This Work$^{\mathrm{c,d,e}}$ & 538 & 59562.471 $\pm$ 0.182 \\
This Work$^{\mathrm{c,d,e}}$ & 539 & 59598.881 $\pm$ 0.451 \\
This Work$^{\mathrm{c,d,e}}$ & 540 & 59632.483 $\pm$ 0.057 \\
This Work$^{\mathrm{c,d,e}}$ & 541 & 59667.256 $\pm$ 0.102 \\
This Work$^{\mathrm{c,d,e}}$ & 542 & 59701.000 $\pm$ 0.070 \\
This Work$^{\mathrm{c,d,e}}$ & 543 & 59736.367 $\pm$ 0.058 \\
This Work$^{\mathrm{c,d,e}}$ & 544 & 59768.933 $\pm$ 0.089 \\
This Work$^{\mathrm{c,d,e}}$ & 545 & 59804.320 $\pm$ 0.061 \\
This Work$^{\mathrm{c,d,e}}$ & 546 & 59839.971 $\pm$ 0.155 \\
This Work$^{\mathrm{c,d,e}}$ & 547 & 59874.247 $\pm$ 0.061 \\
This Work$^{\mathrm{c,d,e}}$ & 548 & 59909.153 $\pm$ 0.060 \\
This Work$^{\mathrm{c,d,e}}$ & 549 & 59943.695 $\pm$ 0.060 \\
This Work$^{\mathrm{c,d,e}}$ & 550 & 59978.842 $\pm$ 0.071 \\
This Work$^{\mathrm{c,d,e}}$ & 551 & 60013.255 $\pm$ 0.055 \\
This Work$^{\mathrm{c,d,e}}$ & 552 & 60048.614 $\pm$ 0.665 \\
This Work$^{\mathrm{c,d,e}}$ & 553 & 60082.703 $\pm$ 0.104 \\
This Work$^{\mathrm{c,d,e}}$ & 554 & 60117.599 $\pm$ 0.117 \\
This Work$^{\mathrm{c,d,e}}$ & 555 & 60152.505 $\pm$ 0.127 \\
This Work$^{\mathrm{c,d,e}}$ & 556 & 60185.828 $\pm$ 0.179 \\
This Work$^{\mathrm{c,d,e}}$ & 557 & 60220.351 $\pm$ 0.057 \\
This Work$^{\mathrm{c,d}}$ & 558 & 60255.629 $\pm$ 0.073 \\
This Work$^{\mathrm{c,d,e}}$ & 559 & 60289.609 $\pm$ 0.060 \\
This Work$^{\mathrm{c,d,e}}$ & 560 & 60322.717 $\pm$ 0.087 \\
This Work$^{\mathrm{c,d,e}}$ & 561 & 60357.549 $\pm$ 0.307 \\
This Work$^{\mathrm{c,d,e}}$ & 562 & 60392.120 $\pm$ 0.075 \\
This Work$^{\mathrm{c,d,e}}$ & 563 & 60427.803 $\pm$ 0.061 \\
This Work$^{\mathrm{c,d,e}}$ & 564 & 60462.501 $\pm$ 0.088 \\
This Work$^{\mathrm{c,d,e}}$ & 565 & 60497.984 $\pm$ 0.058 \\
This Work$^{\mathrm{c,d,e}}$ & 566 & 60533.171 $\pm$ 0.159 \\
This Work$^{\mathrm{c,d,e}}$ & 567 & 60566.725 $\pm$ 0.086 \\
This Work$^{\mathrm{c,d,e}}$ & 568 & 60601.944 $\pm$ 0.053 \\
This Work$^{\mathrm{c,d,e}}$ & 569 & 60636.367 $\pm$ 0.080 \\
This Work$^{\mathrm{c,d,e}}$ & 570 & 60670.166 $\pm$ 0.158 \\
This Work$^{\mathrm{c,d}}$ & 571 & 60704.699 $\pm$ 0.071 \\
\hline 
\end{tabular}
\end{minipage}

\begin{minipage}[t]{0.34\textwidth}
\begin{tabular}[t]{|lcc|}
\hline

\hline 
\end{tabular}
\end{minipage}
}
\end{table*}

\end{document}